\documentclass[aps,prl,twocolumn]{revtex4-2}
\usepackage[T1]{fontenc} 
\usepackage{times}
\usepackage{graphicx}
\usepackage{amsmath}
\usepackage{amsfonts}
\usepackage{amssymb}
\usepackage{multirow}
\usepackage{graphicx}
\usepackage{color}
\usepackage{array}
\usepackage{mathrsfs}
\begin{document}
\title{Translational Dynamics of Rod-like Particles in Supercooled Liquids : 
Probing Dynamic Heterogeneity and Amorphous Order}
\author{Anoop Mutneja}
\email{anoopm@tifrh.res.in}
\author{Smarajit Karmakar}
\email{smarajit@tifrh.res.in}
\affiliation{Tata Institute of Fundamental Research,
36/P,Gopanpally Village, Serilingampally Mandal,Ranga Reddy District, 
Hyderabad, 500107, India} 
\date{\today}

       
\begin{abstract}
{The use of probe molecules to extract the local dynamical and structural properties of complex dynamical systems is an age-old technique both in simulations and experiments. A lot of important information which is not immediately accessible from the bulk measurements can be accessed via these local measurements. Still, a detailed understanding of how a probe particle's dynamics are affected by the surrounding liquid medium is not very well understood, especially in the supercooled temperature regime. This work shows how translational dynamics of a rod-like particle immersed in a supercooled liquid can give us information on the growth of the correlation length scale associated with dynamical heterogeneity and the multi-body static correlations in the medium. A unified scaling theory rationalizes
all the observed results leading to the development of a novel yet simple method that is accessible in experiments 
to measure the growth of these important length-scales in molecular glass-forming liquids.}

\end{abstract}
\maketitle
\textit{Introduction:} 
Supercooled liquids are structurally almost indistinguishable from their high-temperature liquid state 
structures as probed using various scattering experiments. Even with a significant decrease in temperature, 
they remain so unlike the normal crystallization phenomenon where continuous translational symmetry is 
broken and crystalline order sets in at freezing transition leading to the appearance of rigidity. The similar 
appearance of rigidity in supercooled liquids while approaching the so-called calorimetric glass transition 
temperature manifests itself in the massive increase in the viscosity (or structural relaxation time) of the system \cite{DebNat} 
with no emerging apparent structural order, which makes it extremely hard to rationalize \cite{Dyre}. But on the other 
hand, the individual particles of supercooled liquids are found to be dynamically correlated to each other over 
a length scale of $\xi_d(T)$ that grows with increasing supercooling \cite{Dyre,HeteroRev,KDSAnnualReview,Berthier2011}. This knowledge portrays a picture of the 
supercooled liquid state as a medium of different spatial regions with very different dynamical properties, 
often given the name of Dynamical Heterogeneity (DH) in the literature\cite{Berthier2011,HeteroRev,Keys,KDSAnnualReview}. The framework of DH can very 
efficiently explain the experimentally observed features like stretched exponential relaxation, decoupling of 
viscosity and diffusion constant, non-Gaussian self part of van Hove function, rotation-translation decoupling \cite{HeteroRev,Blackburn1996,WeeksExpt}, etc. Previous experiments on 
colloidal glasses also confirm the growth of $\xi_d(T)$ \cite{ColloidWeeks,Hima2015,Gokhale2014}. However, its experimental measurements in molecular 
liquids like glycerol, \textit{ortho}-terphenyl (OTP), sorbitol, etc., remain a challenge. Only a handful of complicated 
procedures exist if one wants to measure the same in experiments using higher order correlation functions\cite{Berthier2005}.
\begin{figure}[!ht]
\vskip -0.1in
\includegraphics[width=0.49\textwidth]{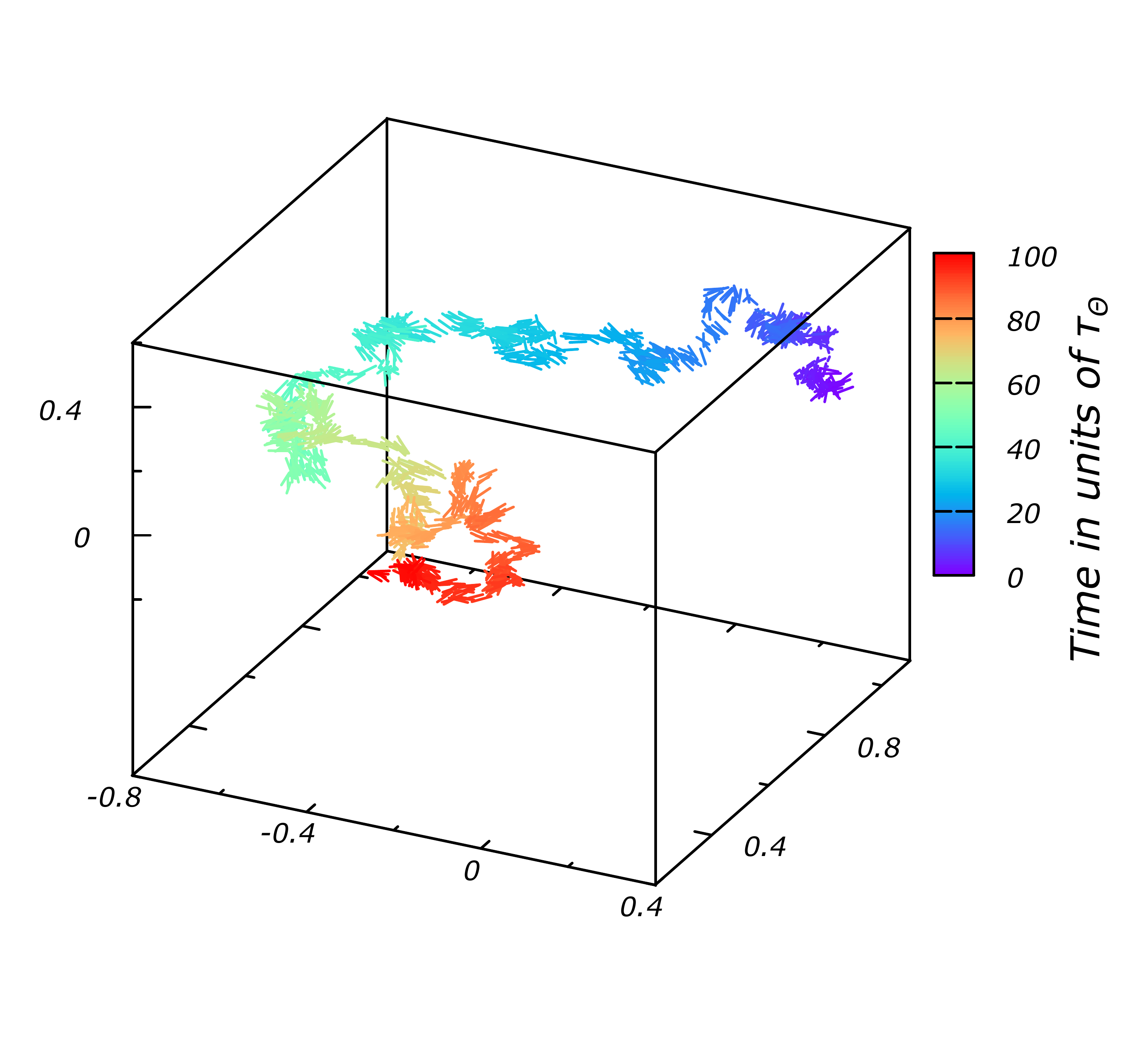}
\vskip -0.35in
\caption{The rod's trajectory in a model supercooled liquid at low temperature. The color bar represents the rod's time evolution in the units if $\tau_\theta$, where $\tau_\theta$ is the averaged time taken for a rod to rotate by unit radian. A cage motion with intermittent jumps clearly signifies the underlying heterogeneity in the embedding supercooled liquid medium's dynamics.}
\label{Fig1}
\end{figure}

On the other hand, as suggested in previous works \cite{SKPNAS,KDSAnnualReview,KDSROPP}, the growing, dynamical correlations seem to not causally 
account for the rapid growth of liquid's viscosity or structural relaxation time.  There has to be an 
accompanying increase of some static correlation often termed in the literature as ``Amorphous order"~\cite{AG,Kauzmann}. 
This growth of correlation amongst disorder manifests itself in the observed finite-size effects of the lowest 
vibrational frequency of the dynamical (Hessian) matrix \cite{SKEigen}, of structural relaxation time $\tau_
\alpha$ (defined later) \cite{SKPNAS}, etc. Thus, this correlation can be measured via careful finite-size scaling (FSS) of 
these quantities \cite{SKEigen,SKPNAS,KDSAnnualReview}. This length scale is also related to configurational entropy $S_c$ proposed in Adam-Gibbs 
relation \cite{AG}, hence giving other ways to quantify it in the simulations \cite{SKPNAS}. Alternatively, this 
static correlation can also be measured using the so-called point-to-set (PTS) correlation function \cite{PTS}. Still, this method will apply only to simulations and colloidal experiments due to its complex procedure. On the other hand, for molecular liquids, the existing experimental measurements for growing static correlation
are extremely subtle and are done by measuring the fifth-order dielectric susceptibility $\chi_5(t)$ in supercooled 
glycerol and propylene \cite{BiroliPaper}. Thus it won't be incorrect to say that the present-day experiments 
are not equipped to quantify both static and dynamic length scale in molecular supercooled liquids easily. Any 
new proposal for the experimental method to quantify both these length-scales would then be highly welcomed. 
Future insights in growth of these two length scales would have far-reaching implications in developing a unified 
theory of glass transition and industrial applications like bio-preservation \cite{Bio1,Bio2,Bio3,Bio4,Bio5}.

The new method proposed in this letter can also lead to a better understanding of the existing results of 
single-molecule (SM) experiments  and, at the same time, help us simultaneously extract the important length 
scales of the system. In the current SM experimental results, the dynamical heterogeneity is quantified either 
via decoupling of rotational and translational motions of probe molecules or by the observed broadening in 
the rotational relaxation time distribution, $\tau_\theta$ (see SI for definition) 
\cite{HeteroRev,Blackburn1996,WeeksExpt,KaufmanRev}. The observed decoupling 
between rotational and translational motions of elongated probe molecules can be very well rationalized by 
dynamical heterogeneity in the medium. On the other hand, the observed broadening in the rotation 
relaxation time, although claimed to be a direct implication of dynamical heterogeneity, is related to the 
growing static length scale of amorphous order. In a recent work \cite{Anoop}, the exact form of the 
distribution of rotational relaxation time has been derived in the high-temperature limit, and deviation from this 
asymptotic form is then used to quantitatively measure the growth of static length scale via an elegant scaling 
analysis. Similarly, dynamic length scale $\xi_D$ is shown to be obtained from the scaling analysis of the 
non-normal parameter \cite{Jain}, which measures the deviation in rotational displacement 
distribution from the usual high-temperature homogeneous behaviour. In this letter, we have done a detailed 
study of the dynamical effects induced by the supercooled liquid environment on the translational dynamics of 
the center of mass (CoM) of a probe particle with an increasing degree of supercooling and for different probe 
sizes.

The main idea behind this work is as follows, on spatial coarse-graining of the supercooled medium over a 
length scale similar or larger than the underlying correlation length, the system will start to behave like a 
homogeneous liquid medium \cite{Block,BhanuPRE,Anoop}. This coarse-graining procedure unfortunately 
would require one to have the positions of all the constituent particles in the system across various timescales, but 
this will be nearly impossible to implement for molecular liquids. One can avoid this difficulty by obtaining the system's 
response averaged over the desired length scale via a probe molecule. This is precisely the idea used in this work 
to extract the liquid medium's response at various length scales by systematically changing the probe size 
similar to the SM experiments. The SM probes of different linear lengths would effectively probe the system at 
the respective coarse-graining length and significantly reduce the experimental complexity. The only task at 
hand would be to appropriately choose the probes of various sizes for a given supercooled liquid and study 
their dynamics as a function of the varying probe length. In Ref.\cite{Anoop}, the validity of this concept has 
been rigorously established for the rod's rotational dynamics (probe) molecule. In this letter, we show how 
dynamical and structural information can be obtained similarly using the translational motion of the CoM of a 
rod with the possibility of lesser experimental complexity.


\begin{figure}[!ht]
\includegraphics[width=0.48\textwidth]{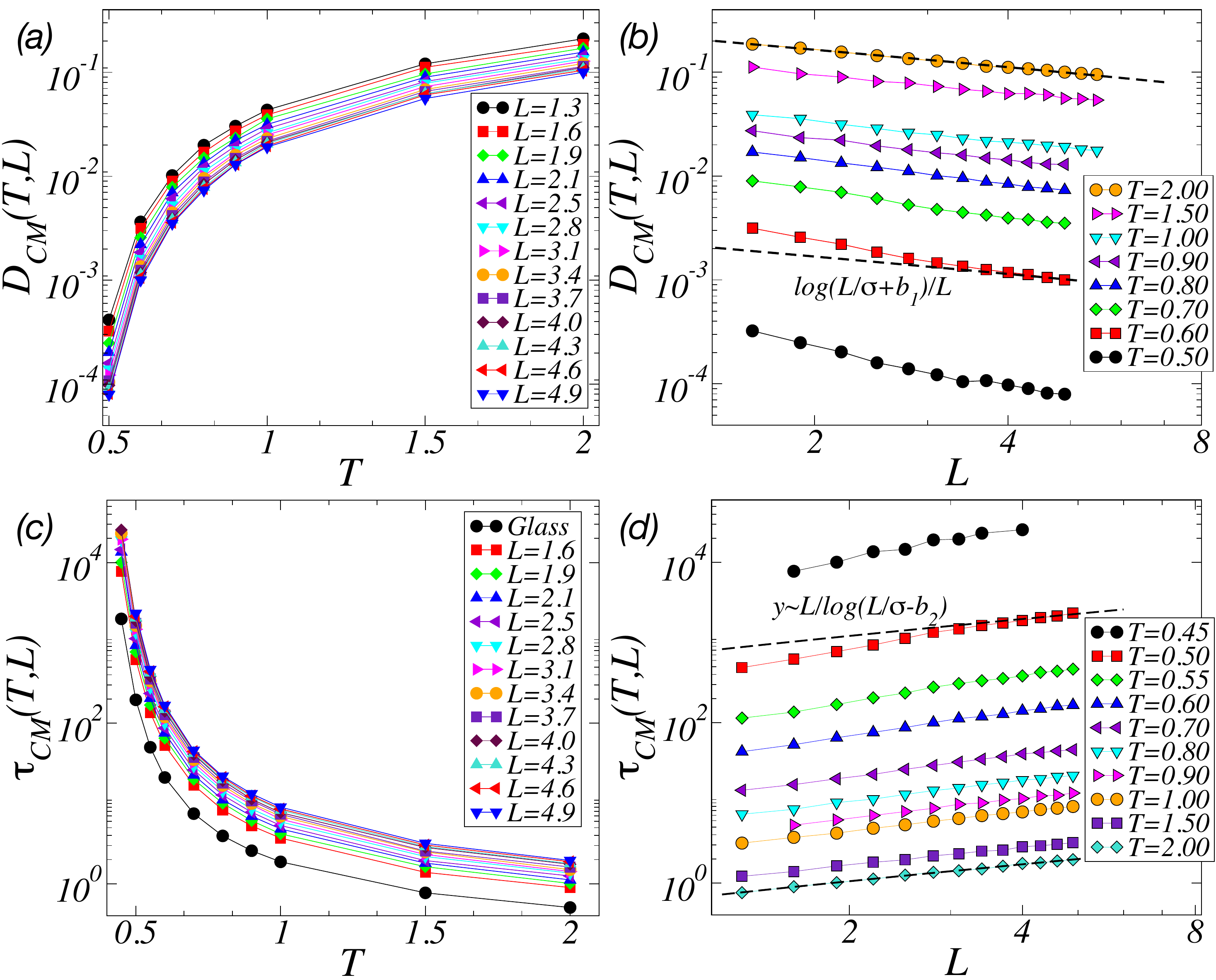}
\caption{Left Panels: Temperature variation of the diffusion constant, $D_{CM}$ (top) and the relaxation time, $\tau
_{CM}$ (bottom) of CoM of the rod of different lengths immersed in 3dKA medium. Right Panels: Rod-length variation of $D_{CM}$ and $\tau_{CM}$ for rods immersed in 3dKA system at different temperatures. The rod-length variation for small rod lengths and high-temperature regimes matches the Brownian particle's predicted hydrodynamic expression (dotted lines).}
\label{DCMTauCM}
\end{figure}

\textit{ Models and Methods: }
This work contains extensive simulations of two model glass-forming liquids, namely - (i) the Kob-Andersen 
$80:20$ binary Lennard-Jones mixture (3dKA) \cite{KA} and (ii) the $50:50$ binary mixture of particles 
interacting via harmonic potential (3dHP) \cite{HP} in three dimensions. We have not looked at models in two 
dimensions primarily because of long-wavelength fluctuations arising due to the Mermin-Wagner theorem, 
which affects quantities related to particles' translational motion in two dimensions \cite{MW,Cage1}. Although these 
effects can be taken care of systematically by computing appropriate correlation functions \cite{Cage1,Mazoyer_2009,SaurishCage,Li2019}, thus the 
results presented in this paper are not limited to three-dimensional systems only. The units of length, energy, and time are the diameter of the larger particle ($\sigma$), the pre-factor of the potential energy function ($
\epsilon$), and ($\sqrt{m \sigma^2/\epsilon}$), respectively. Here $m$ is the mass of a constituent 
particle. The other reduced units can be derived from these 
units. Further details of the models, and the techniques used 
to perform the constant number of particle ($N$), pressure ($P$) and temperature $T$ molecular dynamics simulations (NPT) with few rod-like particles are given in the SI. We want to 
highlight that we have chosen these two model systems in this study primarily to understand the generic nature 
of the presented results. The 3dKA model is developed to model the supercooled liquid dynamics of 
molecular glass-forming liquids. The 3dHP model is a paradigmatic model to study jamming physics in the 
context of dense, soft sphere systems and soft granular particles, including colloidal particles in experiments. 
The rest of the letter is broadly divided into two parts. In the first part, we discuss the scaling analysis 
performed to extract the static length scale, $\xi_S$, from the rod length dependence of relaxation time and 
diffusion coefficient. Then, we discuss the Stokes-Einstein (SE) and Stokes-Einstein-Debye (SED) relation in 
the same context and show how the departure from both SE and SED relation can be used to obtain the 
growth of the underlying dynamic length scale, $\xi_D$. It is fascinating to see how all the probe particle's 
dynamical behaviour can be rationalized using these two length scales in the system via an elegant scaling 
analysis.

\begin{figure*}[!htpb]
\includegraphics[width=0.97\textwidth]{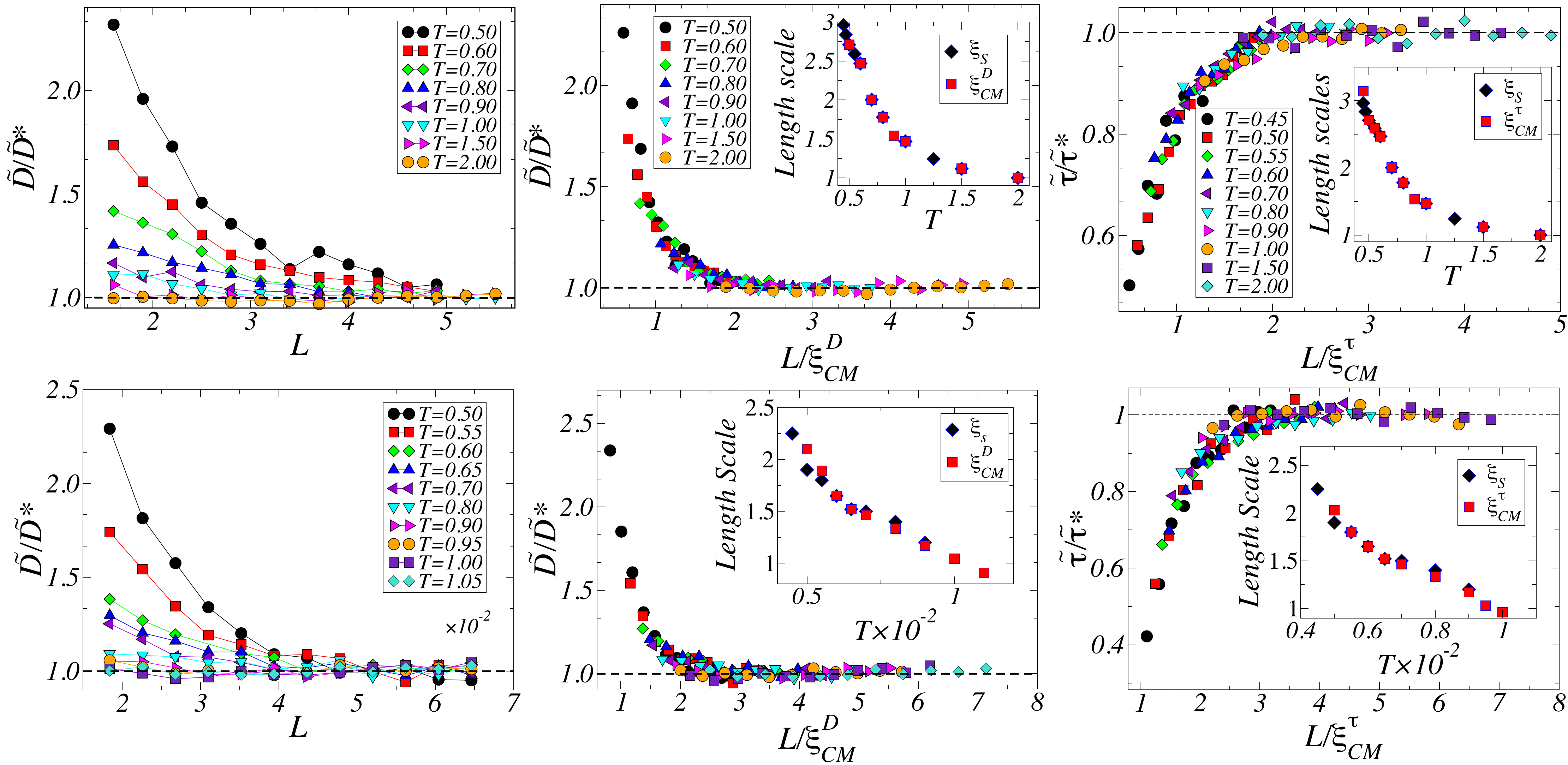}
	\caption{Top-Panel: In the left-panel, $\tilde{D}/\tilde{D}^*$ is plotted against the rod length of rods in different 
	supercooling temperatures of the 3dKA model. Here $\tilde{D}(T,L)$ is the $D_{CM}$ with the rod-length 
	dependence of homogeneous medium removed and $\tilde{D}^*(T)=\tilde{D}(T,L\rightarrow\infty)$. The parameter 
	$\tilde{D}/\tilde{D}^*$ deviates from the homogeneous behaviour (constant in $L$) at larger and larger rod lengths 
	with increasing supercooling (while approaching from large rod-length). Center-panel contains the data collapse of 
	$\tilde{D}/\tilde{D}^*	$ using the scaling ansatz of Eq.\ref{DcmScaling}. The temperature dependence of the 
	obtained length scale is similar to that of the system's static length scale computed using the PTS method 
	(center-inset) \cite{Block}. Right-panel shows the similar data collapse for  $\tilde{\tau}(T,L)/\tilde{\tau}^*(T)$ using 
	Eq.\ref{TaucmScaling} and compares the obtained length scale to the PTS static length scale \cite{Block} in the
	 inset. The bottom-panels have similar plots for the 3dHP model.}
\label{CollapseDcm}
\end{figure*}
\textit{Translational diffusion constant $D_{CM}(T,L)$ \& relaxation time $\tau_{CM}(T,L)$:}
The essential quantities to study for translational dynamics of rods would be the mean squared displacement 
(MSD) and the two-point overlap correlation function ($Q(t)$) of the CoM of the rods. Also, since the rotational 
correlations for large enough rods can survive for substantially long times in supercooled liquids, the MSD of 
the rod along its initial orientation and perpendicular to it would provide us insight into the translational 
anisotropy of probe rods. Further detailed characterizations, including the definitions of MSD and $Q(t)$, are 
given in the SI. The diffusion constant $D_{CM}(T, L)$ and the relaxation time $\tau_{CM}(T, L)$ of the CoM 
motion of the rod can be estimated from MSD and $Q(t)$, respectively (see SI for details). The temperature 
dependence of $D_{CM}$ and $\tau_{CM}$ for different rod lengths and the rod length dependence at 
different supercooling temperatures for the 3dKA model are shown in Fig.\ref{DCMTauCM}. Also, the plots are 
similar for the 3dHP model and can be found in SI. In a homogeneous liquid, the diffusion constant of a 
Brownian rod should decrease with rod length as $D_{CM} \sim \ln{(L/\sigma + 
b_1)}/L$, if one considers the hydrodynamic interaction with the surrounding medium. Here $\sigma$ is the 
rod's width, and $b_1$ is a correction factor arising due to the drag effect coming from the two ends of the rod 
(see Ref.\cite{Bookpoly}  for details). Similarly, relaxation time will increase as 
$\tau_{CM}\sim L/\ln{(L/\sigma + b_2)}$ in the large rod and 
high-temperature limit with a different constant, $b_2$. This dependence can be clearly seen in the right panels of 
Fig.\ref{DCMTauCM}. Thus,  $\ln(L)/L$ like dependence in $D_{CM}$ can be attributed solely to the rod's translational 
properties in a homogeneous liquid medium. Any subsequent deviation from this behaviour at either smaller 
rod length or lower temperature can be thought of as the effect of growing correlations in the medium due to 
supercooling.

\begin{figure*}[!ht]
\includegraphics[width=0.98\textwidth]{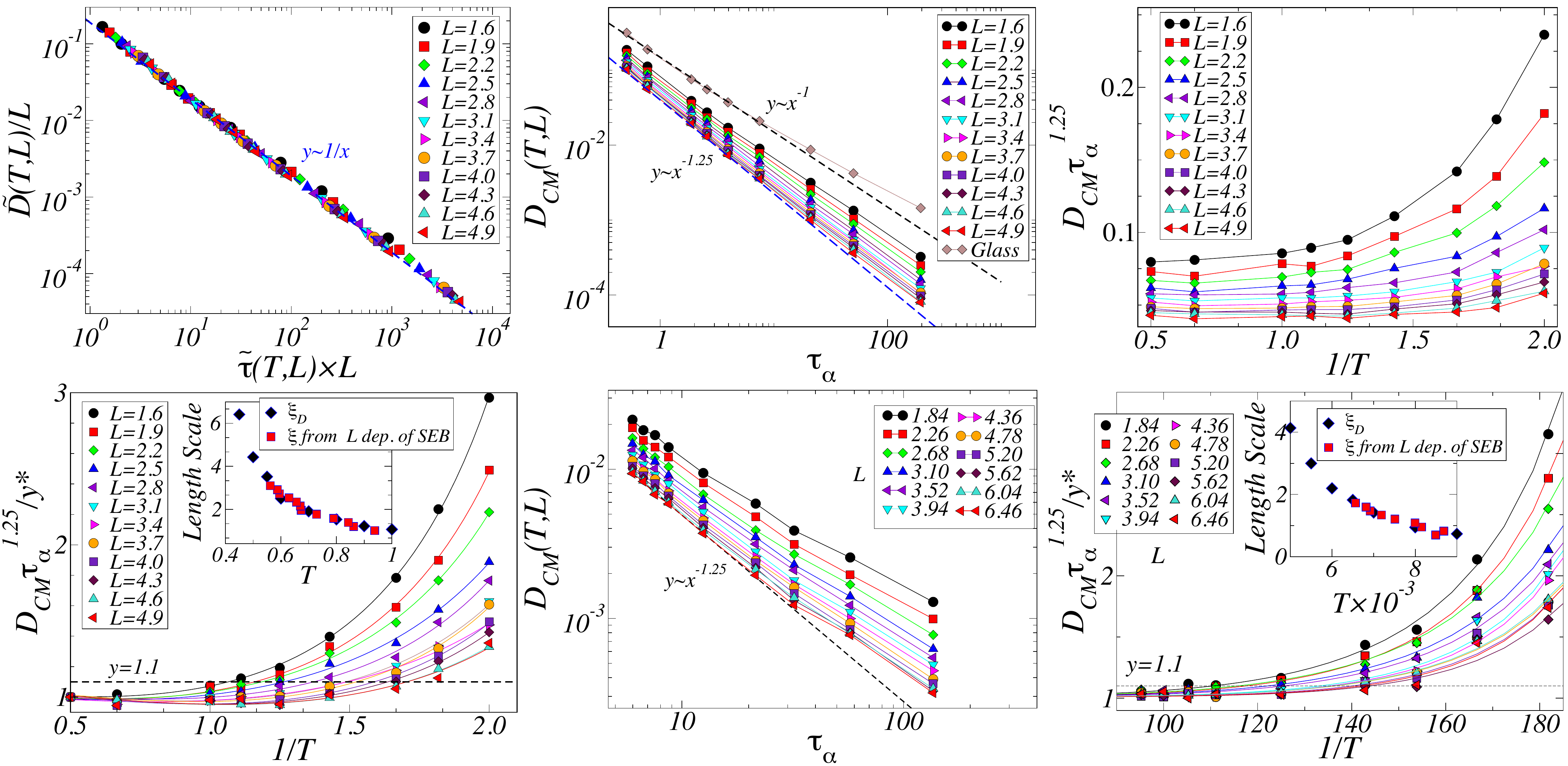}
	\caption{(a.)The cross plot collapse of $\tilde{D}(T,L)/L$ and $\tilde{\tau}(T,L)\times L$ for 3dKA liquid shows the 
	validity of the Stokes-Einstein relation for rods' self variables, along with the same rod-length dependence (even 
	with violations because of supercooling). In SI, one can find a similar plot for the 3dHP model. The cross plot 
	between the $D_{CM}(T,L)$ and the liquids' relaxation time ($\tau_\alpha(T)$) shows the emergent power-law of 
	$D_{CM}\sim \tau_{\alpha}^{-1.25}$ in both 3dKA (b) and 3dHP (e) models for large rod lengths. The fractional SE 
	relation $D_{CM}\tau_{CM}^{1.25}\sim const.~of~T$ is checked and found to be violating at lower and lower 
	temperatures for larger rods in (c) for the 3dKA model. The SEB temperature for various rod lengths is obtained as 
	a $y/y*=1.1$ cut to $y=y*exp[x+c_0(x/x_0)^n+c_1(x/x_0)^{2n}]$ functional fit. Here, $x=1/T$, $y(T,L)=D_{CM}\tau_
	\alpha^{1.25}$, and $y*$, $c_0$, $c_1$, $x_0$, $n$  are the fitting parameters. The procedure is strictly followed as 
	in \cite{}, and is outlined in (d) and (f) panel for 3dKA and 3dHP models. The rod length dependence of SEB 
	temperature is very similar to the system's dynamic length scale and is shown in the respective inset. The missing 
	plots for the 3dHP model can be found in SI.}
\label{SECollapse}
\end{figure*}
Interestingly, both of these quantities start to show violation from logarithmic dependence below a critical rod 
length that increases with increasing supercooling. Physically, both of these violations imply the faster 
translational diffusion of smaller rods in a supercooled environment than it should have been according to the 
underlying homogeneous liquid behaviour. This behaviour is very similar to the observed rotational hops in 
the previous work \cite{Anoop}, making the smaller probes rotationally decorrelate at a shorter time than 
expected. This anomaly was linked with the underlying growing static correlations, so we can expect that the 
growth of the same static length scale can be the cause here. The measured increase (decrease) in $D_{CM}
(T,L)$ ($\tau_{CM}(T,L)$) should then be an effect of this emerging structural order.

To extract the underlying length scale, we first scale-out the asymptotic high-temperature and large rod 
length dependence and look at the parameter $\tilde{D}(T,L)=D_{CM}L/\ln{(L/\sigma + b_1)}$. $\tilde{D}$ is 
then scaled by its large rod-length value at each temperature $T$, i.e., $\tilde{D}^*(T) = \tilde{D}(T,L\to \infty)$. 
In the top left panel of Fig.\ref{CollapseDcm} we show $\tilde{D}(T,L)/\tilde{D}^*(T)$ as a function of rod length 
$L$. It is clear that deviation from high temperature large rod-limit asymptote becomes prominent with 
decreasing temperature, suggesting growth of underlying correlation length in the system. Data for all 
temperature and rod lengths can then be collapsed by plotting $\tilde{D}(T,L)/\tilde{D}^*(T)$ as a function
of $L/\xi_{CM}^D$, where $\xi_{CM}^D$ is the underlying length scale if one assumes the following scaling 
ansatz. 
\begin{equation}
\frac{\tilde{D}(T,L)}{\tilde{D}^*(T)} = \mathcal{F}\left( L/\xi_{CM}^D\right). 
\label{DcmScaling}
\end{equation}
A good data collapse in the middle panel of Fig.\ref{CollapseDcm} clearly demonstrates the validity of the 
scaling assumptions. Similarly, the length scale $\xi_{CM}^\tau$ can be obtained from the 
relaxation time (See top right panel of Fig.\ref{CollapseDcm}), if we assume a similar scaling ansatz as
 \begin{equation}
\frac{\tilde{\tau}(T,L)}{\tilde{\tau}^*(T)} = \mathcal{G}\left( L/\xi_{CM}^\tau\right), 
\label{TaucmScaling}
\end{equation}
where $\tilde{\tau}(T,L) = \tau_{CM}\ln{(L/\sigma + b_2)/L}$ and 
$\tilde{\tau}^*(T) = \tilde{\tau}(T,L\to \infty)$.
The comparison of obtained length scale with the static length scale of liquid is presented in the respective 
insets. The similarity observed in their temperature dependence establishes the robustness of the method. A 
similar analysis for the 3dHP model as shown in the bottom panels of the same figure suggests the method's 
generic nature across various model systems that are very different. Surprisingly, the scaling functions used 
for the $D_{CM}$ and $\tau_{CM}$ scaling analysis (see Eqs. \ref{DcmScaling} and \ref{TaucmScaling}) turn 
out to be inverse of each other, leading to a strong coupling between them. Thus if one plots $\tilde{D}_{CM}$ 
as a function of $\tilde{\tau}_{CM}$, then one observes fantastic data collapse for all temperatures and rod 
lengths as shown in the top left panel of Fig.\ref{SECollapse}. The lovely master curve implies that even in 
low temperatures, $\tilde{D}_{CM}$ and $\tilde{\tau}_{CM}$ obey the Stokes-Einstein (SE) relation $\tilde{D}
_{CM}(L,T)\tilde{\tau}_{CM}(L,T)= constant$ as will be discussed in the subsequent paragraphs.

\textit{Stokes-Einstein Breakdown (SEB):}
Stokes-Einstein (SE) relation is the blend of Einstein's equation for diffusion constant and Stokes's equation 
for drag coefficient for a spherical particle in a homogeneous medium. It reads as $D = k_B T/c\pi R\eta$, 
where $D$ and $R$ respectively are the diffusion constant and the particle's hydrodynamic radius. $\eta$ 
and $T$ are the shear viscosity and the liquid's temperature, respectively, and $c$ is a constant which 
depends on the boundary condition at the particle's surface ($c = 6$ for stick \& $c = 4$ for slip) 
\cite{LandauFD}. $k_B$ is the Boltzmann constant. This expression is also valid for the self-diffusion of the 
liquid particles. Breakdown of this famous Stokes-Einstein relation is believed to be one of the hallmarks 
of dynamic heterogeneity in supercooled glass-forming liquids. Since the distribution of local viscosity or structural 
relaxation time gets broader due to dynamic heterogeneity, the diffusion constant's growth becomes faster 
than the predicted value from SE relation. The viscosity in this expression is often replaced by $\tau \sim \eta/
T$, as the growth of these two quantities is essentially the same over the studied temperature range. Thus 
the Stokes-Einstein relation for a rod-like particle will read as, 
\begin{equation}
D\tau=\frac{1}{c\pi R}=f(R). 
\end{equation}
$f(R)$ can be considered as a measure of SEB. Now, for dynamics of rods in the liquid, one can have 
$D_{CM} \tau_{CM}$ as well as $D_{CM} \tau_\alpha$ as possible SEB parameters. Where $D_{CM}$ \& $
\tau_{CM}$ are the diffusion constant and relaxation time of CoM of the rod and $\tau_\alpha$ is the 
relaxation time of liquid (see SI for definition). These two SEB parameters look at the validity of SE relation 
for the rod's self dynamics and the coupling of the dynamics of the rod with the liquid medium. Thus, it gives 
rise to a possibility of understanding how dynamical information of liquid can be obtained from the rod's 
dynamics in that medium. The top-left panel of Fig.\ref{SECollapse} confirms the validity of the SE relation for 
$D_{CM}$ and $\tau_{CM}$ of the rod particles. This tells us that the SE relation between the rod's diffusion 
constant and relaxation time is obeyed for all temperatures, even deep in the supercooled liquid regime for all 
rod lengths.

Surprisingly, the fractional SE relation $D_{CM} \sim \tau_\alpha^{-\omega}$ seems to be obeyed in the 
large-rod-length and high-temperature limit with an emerging power-law exponent $\omega = 1.25$. Note 
that exponent is bigger than 1, and in supercooled liquids, with diffusion constant and relaxation time of the 
liquid, one sees similar power-law dependence with exponent $\omega$ smaller than 1. Although we do not 
have a microscopic understanding of this exponent's value for rod-like particles, it is interesting to note that 
this exponent seems universal across the two model systems that we have looked at in this work. Note that 
for smaller rod lengths, the power-law dependence appears to be deviating from the asymptotic behaviour, 
which can be quantified if we plot $D_{CM}\tau_\alpha^{\omega}$ as a function of inverse temperature as 
shown in Fig.\ref{SECollapse} (bottom-left) for different rods. As seen, this parameter remains almost 
constant for the largest rod for the entire temperature range, with deviation showing up at the lowest studied 
temperature. The SE breakdown starts to happen at a higher and higher temperature with decreasing rod 
length, as shown in the top right panel of Fig.\ref{SECollapse}. This behaviour is remarkably similar to the 
results reported in Ref.~\cite{SastrySE} for the wave vector ($k$) dependence of SEB. It was found that the 
SEB at various $k$ occurs at lower and lower temperatures as one decreases the value of $k$ (large 
wavelength). In their case, the $k$ dependence of SEB temperature was found directly related to dynamic 
heterogeneity length scale $\xi_D$. One can rationalize the similarity to our system by assuming that the rod 
measures the liquid's response at a length scale comparable to its size. We can then compare it with the 
results obtained at a wave vector $k \sim 1/L$, where $L$ is the rod length. Motivated by this, we also 
repeated the analysis reported in Ref.\cite{SastrySE} to see whether the rod length dependence of SEB 
temperature correlates with the temperature dependence of liquid's DH length scale. 
In Fig.\ref{SECollapse} bottom-right panel, we have shown 
such an analysis. To obtain SEB temperature for different rods, one has to take $1/T$ value at $y/y^*=1.1$, 
where $y(T,L)=D_{CM}\tau_\alpha^{1.25}$ and $y^*(L)=y(2.0,L)$. This procedure is very similar to the 
one adopted in Ref.\cite{SastrySE} and is outlined in the bottom-right panel of Fig.\ref{SECollapse}. The solid 
lines in the figure are the fit to function $ln(y/y*)=[x+c_0(x/x_0)^n+c_1(x/x_0)^{2n}]$ where $x=1/T$. The rod 
length dependence of SEB temperature turns out to be very similar to the system's dynamic length scale, as 
shown in the inset of Fig.\ref{SECollapse} bottom-right. We have shown the results obtained for 3dKA and 
3dHP models in the bottom left and right panels. The generic nature of the results in different model systems 
again highlights the robustness of the proposed method to obtain the dynamic heterogeneity length scale for 
supercooled liquids in simulations. 

\begin{figure}
\includegraphics[width=0.48\textwidth]{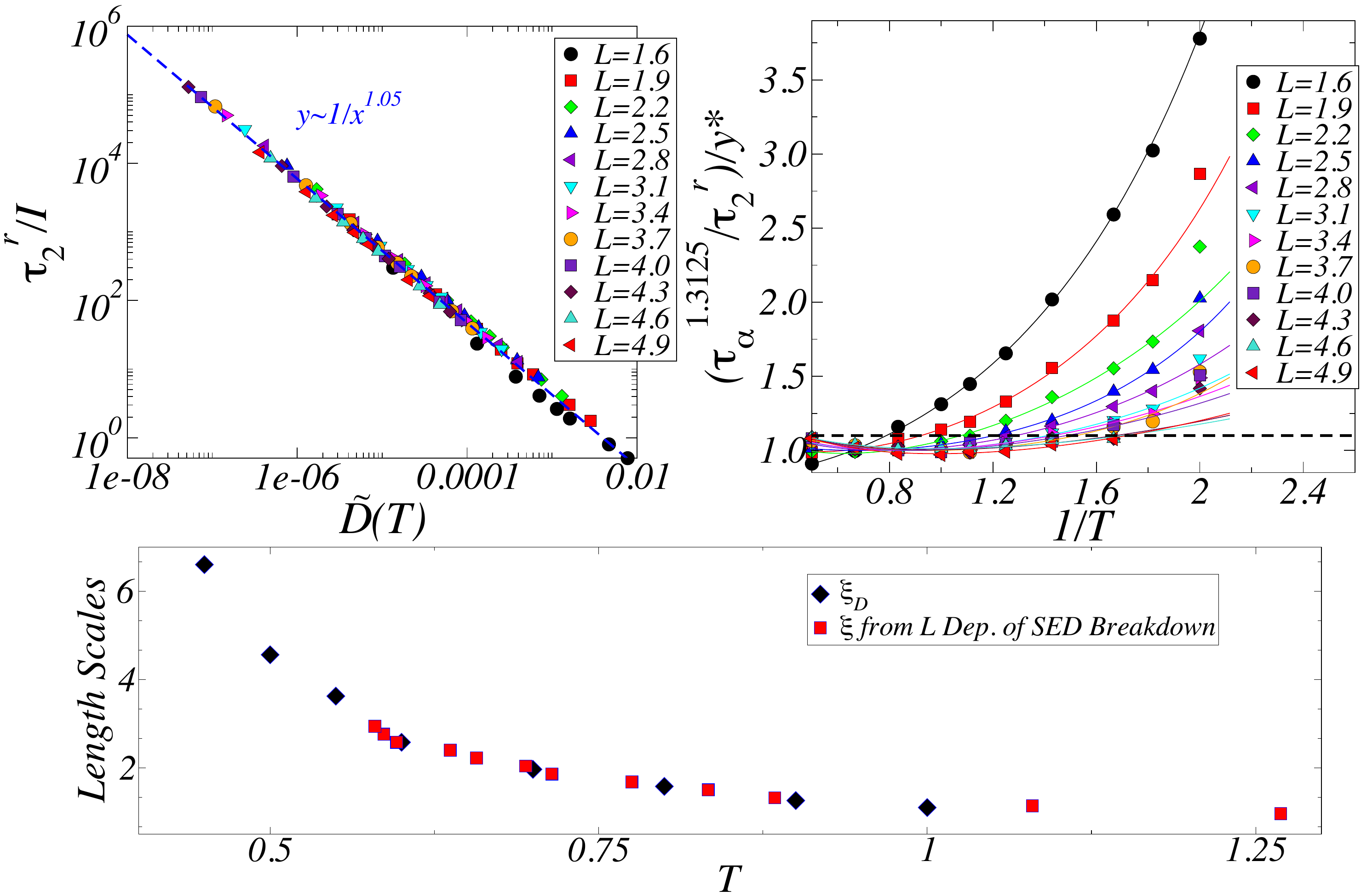}
\caption{(a). The 2nd order rotational relaxation time $\tau_2^r(T,L)$ is plotted against the translational diffusion constant $D_{CM}(T,L)$ to make the one-to-one correspondence between the two. Implying the SEB analysis on $\tau_2^r$ would also provide us the growing dynamical correlations in the system. (b). The SEB analysis is done on variable $\tau_\alpha^{1.3125}/\tau_2^r$ to extract the length scale which is compared with the system' dynamic length scale in (c). This analysis is shown for 3dKA system, similar for the 3dHP model can be found in SI.}
\label{RotTransSE}
\end{figure}

\textit{Breakdown of Stokes-Einstein-Debye (SEDB):}
The rotational dynamics of the probe rod is although independent of translation dynamics but can shape the 
translational dynamics to a large extent. This makes the study of rotation-translation coupling important in the 
homogeneous medium itself. On the other hand, with supercooling, the rotational-translational 
dynamics tend to decouple \cite{HeteroRev,Blackburn1996,WeeksExpt}. Translational diffusion is enhanced compared to the rotational diffusion and the 
medium's viscosity with supercooling. Similar behaviour is observed in our probe-rods. The detailed 
characterization of the same can be found in SI. Also, because of the intimate coupling of the rotational and 
translational properties of probes, one can use them as a proxy to each other. In Fig.\ref{RotTransSE}(a), the 
one-to-one correspondence is made between the rotational relaxation time $\tau_2^r$ (see SI for definition) 
and $D_{CM}$. The fantastic data collapse is obtained when one plot $\tau_2^r/I$ with $I$ being the moment 
of inertia of the rod against $\tilde{D}(T)$ for all the rod lengths and temperatures. The data collapse 
suggests a power-law relation between these two quantities as $\tau_2^r\sim D_{CM}^{-1.05}$, with an 
exponent which is apriori not immediately known to us. This relation also suggests faster translational than 
rotational motions with increasing supercooling. Since in experiments, the viscosity or the relaxation time of the host 
liquid along with the rotational relaxation time of probes are quite easy to get, one can study the breakdown 
of fractional Stokes-Einstein-Debye (SED) to obtain the dynamic length scale as follows. As $\tilde{D}(T) \sim 
\tau_\alpha^{-\omega}$ in the asymptotic high temperature and large rod length limit, we can expect $
\tau_2^r$ to obey similar power-law relation with $\tau_\alpha$ as  $\tau_2^r \sim \tau_\alpha^{\omega_r}$ 
with an exponent $\omega_r  = 1.05\omega \equiv 1.3125$. We need to plot $\tau_\alpha^{1.3125}/\tau_2^r$ as 
a function of $1/T$ to extract the liquid's dynamical length scale in the same manner before. Fig.
\ref{RotTransSE} (b) \& (c) provide the analysis and compare the obtained length scale with $\xi_D$ for the 
3dKA model. A similar analysis for the 3dHP model can be found in SI. The excellent match between the 
length scale obtained via this analysis and the dynamic length scale, $\xi_D$ obtained using conventional 
methods, is remarkably good. This suggests that a complete analysis of Stokes-Einstein and Stokes-Einstein-
Debye relation of a probe particle in supercooled liquid medium can unambiguously extract out the growth of 
underlying dynamic length scale in experimentally relevant glass-forming liquids. 

\textit{Experimental Feasibility:} 
Next, we discuss the experimental feasibility of the proposed method for extracting the dynamic and static length scale in glass-forming liquids. The method presented in this manuscript requires one to record the translational and rotational dynamics of elongated probes of different linear lengths. In Table.~\ref{exptTable}, we have listed
some of the existing experimental works that have already measured various rotational and translational quantities of interest for different sized probe molecules in supercooled liquid media. With these existing experimental results, we believe that our proposed method is feasible in experiments without any additional experimental complexity. In fact, our method proposes a simple but direct way to obtain both the dynamic and static length scales of the supercooled liquids from the same experiment.	
\begin{table}
\begin{tabular}{ |c|l l| }
\hline
\centering
Medium & \multicolumn{2}{c|}{Probe} \\ \hline
	\multirow{3}{*}{Glycerol}	& (a) tbPDI \cite{Mackowiak2011,Zondervan2007} 
										& (b) dpPDI \cite{Mackowiak2011} \\
										& (c) dapPDI \cite{Mackowiak2011} 
										& (d) Rubrene \cite{Mackowiak2009} \\
										& (e) Gold nano-rods \cite{Yuan2013} & \\ \hline
	\multirow{5}{*}{\begin{minipage}{1.5cm}\centering \textit{ortho}-terphenyl (OTP)\end{minipage}} &	(a) Rubrene \cite{Cicerone1995,Cicerone1996} 
										& (b) BPEA \cite{Cicerone1996} \\
										& (c) Tetracene \cite{Cicerone1993, Cicerone1995,Cicerone1996} 
										& (d)  dpPDI\\
										& (e) Anthracene \cite{Cicerone1993,Cicerone1996} 
										& (f)  tbPDI\\ 
										& (g) egPDI
										& (h) pPDI \cite{Leone2013}\\
										& (i) BODIPY268-dye \cite{Paeng2015}& \\
										\hline										
\end{tabular}
\caption{List of probes used to study the dynamics of supercooled Glycerol and OTP in experiments.}
\label{exptTable}
\end{table}

\textit{Results \& Discussions:} 
To summarise, we have devised a method to obtain the growing many-body static correlation in a supercooled liquid medium by studying the probe size dependence of averaged translational diffusion constant and relaxation time of rod-like probes. A detailed scaling analysis very nicely rationalizes all the observed results for various temperatures and rod lengths in a unified manner. Similarly, we showed that by studying the Stokes-Einstein relation between the diffusion constants of rods and the medium's relaxation time, one could elegantly extract the growing, dynamic correlation length. This result corroborates very well
previous simulation studies on the breakdown of Stokes-Einstein relation in supercooled liquid at various probing wave vectors.
Additionally, on considering the self properties of the rod, the Stokes-Einstein relation is found to be obeyed regardless of the rod-length or the degree of supercooling (at least in the studied temperature range), implying the rods' Brownian nature. Next, we expect to have an excellent noise-to-signal ratio from an experimental perspective since the measured quantities are mostly the first moments of various correlation functions. This further translates to the method's usefulness to
simultaneously study the growth of static and dynamic correlation lengths in molecular glass-forming supercooled liquids. This method's feasibility in experiments is also immediately apparent with the availability of probe particles of varying sizes and the existing relevant experimental techniques using these probes, albeit mainly for the rotational motion. Thus, a systematic study of these probe particles' translational motions in supercooled liquids would indeed serve our purpose. We end this article with the note that understanding the growth of these length scales in glass-forming molecular liquids is most important as one observes $13$ to $14$ orders of magnitude of increase in the viscosity or relaxation time while approaching glass transition, unlike in colloidal glasses or in computer simulations where only first $3$ to $4$ decades of slowing down can be studied.

\textit{Acknowledgement: }
This project is funded by intramural funds at TIFR Hyderabad from the
Department of Atomic Energy (DAE). Support from Swarna Jayanti
Fellowship grants DST/SJF/PSA-01/2018-19 and SB/SFJ/2019-20/05 are
also acknowledged.

\bibliographystyle{achemso.bst}
\bibliography{Main}
\end{document}


\title{Translational Dynamics of Rod-like Particles in Supercooled Liquids : 
Probing Dynamic Heterogeneity and Amorphous Order- Supplemental Material }
\author{Anoop Mutneja}
\email{anoopm@tifrh.res.in}
\author{Smarajit Karmakar}
\email{smarajit@tifrh.res.in}
\affiliation{
Tata Institute of Fundamental Research, 
36/P, Gopanpally Village, Serilingampally Mandal,Ranga Reddy District, 
Hyderabad, 500107, Telangana, India }

\maketitle
\tableofcontents
\section{Definitions/Methods}
\subsection{van Hove Function}
The van Hove distribution function, $G(\textbf{r},t)$, is the dynamical correlation 
function. It provides us the correlation between two particles separated by distance 
$r$ and time $t$. It is defined as the probability density of particle `$i$' at position `$
\textbf{r}$' and time `$t$', with the prior knowledge of particle `$j$' being at the origin 
at an initial time.   
\begin{equation}
G(\textbf{r},t)=\frac{1}{N}\left\langle\sum_{i=1}^N\sum_{j=1}^N\delta 
\left(\textbf{r}-(\textbf{r}_i (t)-\textbf{r}_j(0)\right)\right\rangle
\label{vanHove}
\end{equation}
The self part of the van Hove function, $G_s(\textbf{r},t)$, is the most commonly 
used quantity and is defined as the probability of finding a particular particle at 
position `$\textbf{r}$' and time `$t$' when it is known to be at origin initially. This can also
be simply thought of as the distribution of the displacements of the particles at a given
time $t$. It is formally defined as 
\begin{equation}
G_s(\textbf{r},t)=\frac{1}{N}\left\langle\sum_{i=1}^N\delta \left(\textbf{r}-
(\textbf{r}_i (t)-\textbf{r}_i(0)\right)\right\rangle.
\label{vanHoveSelf}
\end{equation}
For a Brownian particle, the self part is thus given by the following expression, which 
would also be valid for particles in any homogeneous diffusive system at a large 
enough distance and long enough timescales,
\begin{equation}
G_s( \textbf{r},t)=\frac{1}{(4\pi Dt)^{3/2}}exp\left(-\frac{r^2}{4Dt}\right).
\end{equation}
Here $D$ is the diffusion constant of the system. Thus for such a Gaussian process 
the measure of dynamical heterogeneity in a disordered system will simply be the 
non-Gaussianity of $G_s(\textbf{r},t)$ at that time, implying the Binder cumulant, $
\alpha_2(t)$. The $\alpha_2(t)$ is defined as 
\begin{equation}
\alpha_2(t)=1-	\frac{\left\langle|\textbf{r}|^4\right\rangle}{3\left\langle|
\textbf{r}|^2\right\rangle^2}.
\label{binder}
\end{equation}
The dynamical correlation between different particles is best measured by the 
distinct part of van Hove function and is defined by considering different $i$ and $j$ 
particles in Eq.\ref{vanHove}.

\subsection{Overlap correlation function $Q(t)$ \& Structural relaxation time $\tau_\alpha$ }
\begin{figure}[!h]
\includegraphics[width=0.98\textwidth]{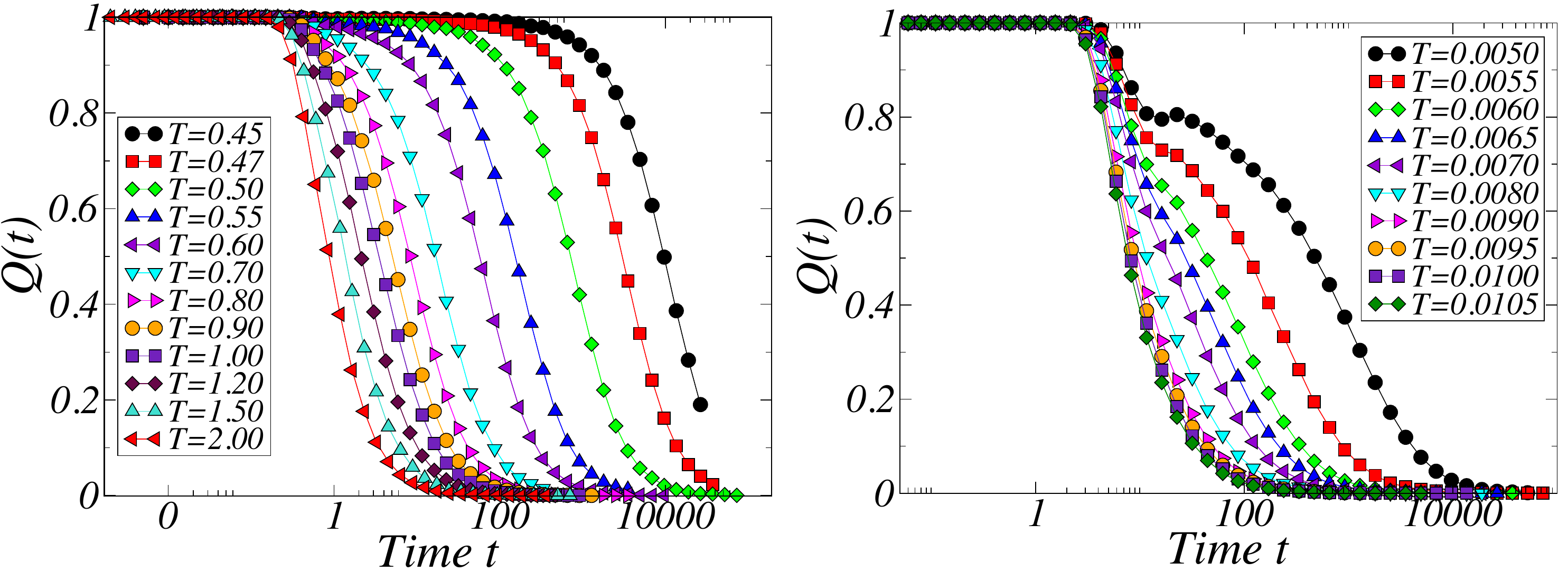}
	\caption{Left-Panel: The overlap correlation function, $Q(t)$ for the rod of length 
	$L=2.5$ immersed in the 3dKA model liquid at different temperatures.
	Right-Panel: The time evolution of $Q(t)$	for the rod of length $L=3.1$ immersed in the 
	3dHP 
	model liquid at different temperatures.
	}
\label{Qt}
\end{figure}

The overlap function, $Q(t)$ gives the measure of overlap between two configurations 
separated by time interval `$t$' and is defined as follows:
\begin{equation}
Q(t)=\frac{1}{N}\sum_{i=1}^N\theta\left(a-|\textbf{r}_i(t)-\textbf{r}_i(0)|\right)
\label{overlap}
\end{equation}
Here, $\theta(x)$ is the usual step function, and the value of $a$ is chosen to ignore 
the decorrelation that might happen due to particles' vibrational motion in their 
cages. In Fig.\ref{Qt}, we have plotted the $Q(t)$ for the rod of length $L=2.5$ \& 
$L=3.1$ immersed in 3dKA (left-panel) \& 3dHP (right-panel) system, respectively, 
at different temperatures. The two step relaxation can be very clearly seen from the 
figures.
The structural relaxation time $\tau_\alpha$ is defined in terms of overlap function. 
It is the time when the average value of the overlap function reduces to $1/e$, i.e., $
\langle Q(\tau_\alpha)\rangle =1/e$. Note that the temperature dependence of $
\tau_\alpha$ doesn't depend on the particular choice of the parameter $a$. Parameter
$a$ is chosen to remove any possible decorrelation that might come due to vibrations
of these particles in their respective cages formed by the neighbours. 

\subsection{Four-Point Susceptibility $\chi_4(t)$} 
Four-Point Susceptibility, $\chi_4(t)$ \cite{Dasgupta1991,Book1} is calculated as the
variance or fluctuation of two point density correlation function 
or the overlap function, $Q(t)$. The variance of overlap function 
 at time $t$ will be the measure of fluctuation in relaxation
in the system at that particular time due to dynamic heterogeneity. It
is defined as    
\begin{equation}
\chi_4(t)=N \left[\left\langle Q^2(t) \right\rangle-\left\langle Q(t) 
\right\rangle^2\right].
\end{equation}
At short times $Q(t)$ is very close to $1$ and goes to zero at larger times so one expects
that $\chi_4(t)$ will be small (close to zero) at very short and long times thus there will be a 
peak at some intermediate timescales. This timescales is found to be very close to the 
$\alpha$-relaxation time, $\tau_\alpha$. This suggests that heterogeneity is maximum 
at the structural relaxation time.  
 
\subsection{Point-to-Set Method}
The causal relation between the growing viscosity of the supercooled liquid and 
the growing length scale of the mosaic state has been advocated in the 
RFOT theory. The more physical way to understand this length scale
is in terms of Point-to-set (PTS) correlation function. One can gain the theoretical 
perspective about PTS length scale, $\xi_{PTS}$ in Ref.\cite{PTS}. 
PTS length scale is calculated by 
taking a well equilibrated configuration of the system and then freezing 
all the particles outside a region $\mathscr{C}$ of linear dimension 
$\xi$ as a boundary condition. The particles inside $\mathscr{C}$ are 
then relaxed and overlap function with the initial configuration is calculated. 
For small enough $\xi$ overlap should be very high implying that the 
particles inside can remain restricted to the same equilibrium state. But 
on increasing the $\xi$ the overlap would start to decrease and vanish, eventually.
Physically, it is because of the fact that this large enough volume can not hold itself 
to single metastable state. 
The largest linear length scale $\xi=\xi_{PTS}$ to which volume of particles can be 
restricted to same metastable minimum is known as PTS length scale. 
This means that the mosaic of this linear dimension can survive in 
particular metastable state, and thus whole system can be thought of 
as divided into such mosaics of metastable states.
The length scale obtained in this way is found to be same as static length scale 
obtained by finite size scaling of $\tau_{\alpha}$\citep{SKPNAS} and from minimum 
eigenvalue of Hessian matrix \cite{SKPhysA}.

\section{Translational of center of mass (CoM) of probe rods in supercooled liquids:}
Anisotropy in the translation of CoM of rod in the homogeneous liquid is evident as 
a rod tends to have a larger diffusion constant in the direction of its orientation ($D_
\parallel$) than any of the perpendicular direction ($D_\perp$). But eventually, with 
rotational decorrelation, this anisotropy disappears with time. The translation of 
CoM of rods is not a Gaussian process because of the strong rotation-translation 
coupling. It becomes Gaussian when the relevant length scale (diffusion 
length) becomes much larger than rod length. Readers are encouraged to read 
chapter-8 of the book \cite{BookPoly} for more details. This section will discuss 
effects of supercooling and rod length on various translational quantities of probe rods. 
The first things that will be discussed are the mean squared displacement (MSD), anisotropy 
observed in diffusion along and perpendicular to the rod, translational overlap 
function $Q_{CM}(t,T,L)$, and the relaxation time associated, etc.

\subsection{MSD of CoM of probe rods}
\begin{figure*}[!h]	
\centering
\includegraphics[width=1\textwidth]{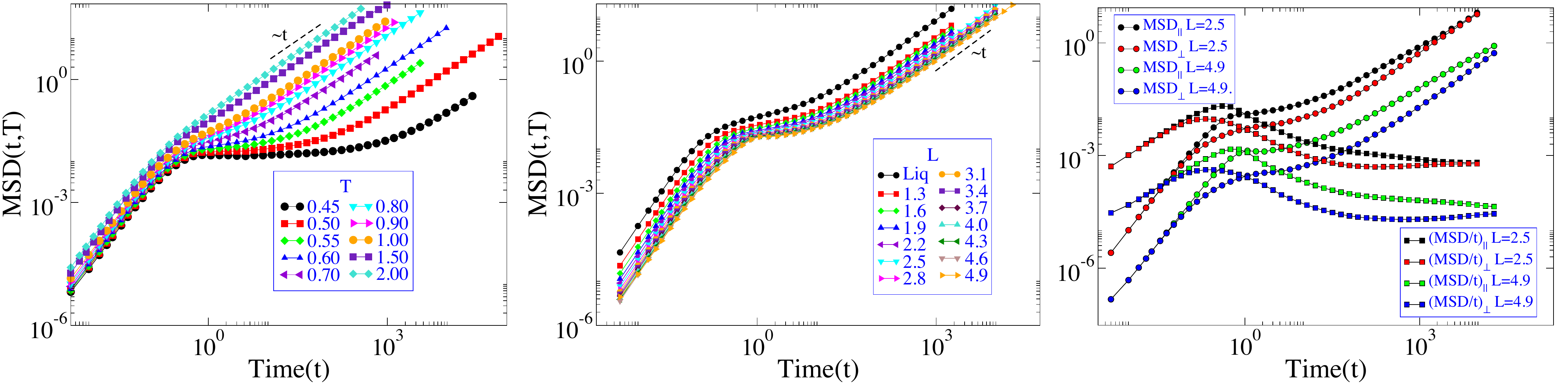}
\caption{Left panel: The time evolution of MSD of rod's CoM with $L=2.5$ immersed 
in 3dKA medium at various supercooling temperatures. 
Center panel: The MSD evolution plot for different sized probes in 3dKA liquid at 
temperature $T=0.6$, for comparison with the parent liquid, MSD of liquid 
particles is also included in the plot. 
Right panel: Evolution of parallel and perpendicular components of 
MSD of rod (circles) for small ($L = 2.5$) and large rods ($L = 4.9$) and the 
corresponding $MSD/t$ (square symbols) components which in long time limit 
are proportional to the diffusion constant. The more diffusion along the rod can 
be clearly seen at the initial and intermediate times but as the rod 
rotationally decorrelates at long times, the MSD along both directions coincides (see text for more discussion).}
\label{TransMsd}
\end{figure*}

The mean squared displacement (MSD) of CoM of the rod is defined as, 
\begin{equation}
MSD=\langle|\textbf{r}_{CM}(t)-\textbf{r}_{CM}(0)|^2\rangle=\langle|\Delta\textbf{r}
_{CM}|^2\rangle
\end{equation}
Where  $\textbf{r}_{CM}(t)$ is the CoM position vector at a time `$t$'. Fig.\ref{TransMsd} 
(left panel) is the time evolution of MSD for the rod of fixed length 
($L=2.5$) immersed in a liquid (3dKA model) at various temperatures in the 
supercooled regime. In this figure's center panel, the rod length variation of MSD is 
shown for probe rod in 3dKA liquid at $T=0.60$. For comparison, the MSD of parent 
liquid particles is also included in this plot.

Following are few remarks one can make from these plots. Firstly, rod's caging duration increases with the supercooling, as expected. Secondly, the larger rods diffuse more slowly than, the smaller ones and the parent liquid particles. The caging duration doesn't change much with the increase in rod length at a given temperature which can be rationalized easily, at least for the diffusion parallel to the rod length since it will depend mainly on the crowding near the tip of the rod. Thus, it will be mostly a function of the medium's density. Also, since the rotational correlations for large enough rods can survive for substantially long times in supercooled liquids, the MSD of CoM of the rod along the rod's initial orientation and the components perpendicular to it, would provide us insight into the translational anisotropy of the probe rods. Fig.\ref{TransMsd} (right panel) contains the plots for the time evolution of the MSD (circle symbols) component parallel to the initial orientation and the orthogonal component. These plots are for rod of size $L = 2.5$ (small) and $L = 4.9$ (large) immersed in 3dKA liquid at $T=0.60$ temperature. One can see that the ballistic time duration lasts longer for the parallel component, which can be justified from geometrical arguments. Also, at large times, both the components should diffuse equally because of the rotational decorrelation, which can be easily speculated. 

In contrast, during the intermediate times, the diffusion along the rod length is larger, consistent with the expected behaviour \cite{BookPoly}. The square symbols in Fig.\ref{TransMsd} (right-panel) represent the MSD per unit time which is linked to the diffusion constant. One can see that the parallel component diffuses faster but eventually would come around to other components as the rod relaxes rotationally. The parallel and perpendicular part of MSD is defined as,

\begin{equation}
MSD_{||}=\langle|\Delta\textbf{r}_{CM}.\hat{\textbf{u}_0}|^2\rangle
\end{equation}             
\begin{equation}
MSD_{\perp}=0.5\langle|\Delta\textbf{r}_{CM}|^2-|\Delta\textbf{r}_{CM}.
\hat{\textbf{u}_0}|^2\rangle,
\end{equation}
where $\hat{\textbf{u}_0}$ is the orientation vector of the rod at time $t=0$. 
There is a half multiplied to the perpendicular component to account for two 
independent orthogonal directions. 

\textit{Rod-length dependence of $MSD_{||}$ and $MSD_{\perp}$:}
\begin{figure}[h]
\includegraphics[width=.8\textwidth]{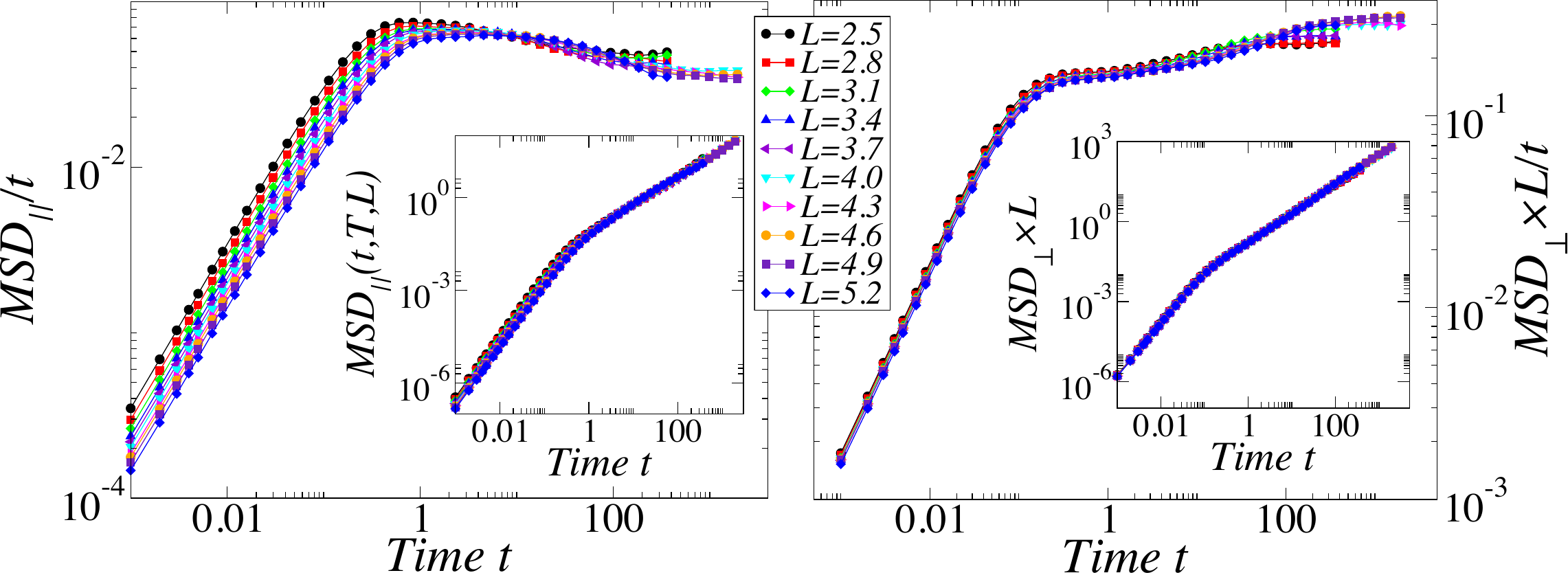}
\caption{Left Panel: The time evolution of $MSD_{||}(t)$ per unit time is shown, along with the $MSD_{||}(t)$ itself in the inset, for rods of various lengths in 3dKA medium at $T=2.0$ (high-temperature). The plot serves the purpose of illustrating the no rod length dependence on the parallel diffusion of rod, until the rotational decorrelation kicks in. Right Panel:  The similar plot of the $MSD_\perp\times L$ per unit time illustrates the $1/L$ rod-length scaling of the perpendicular component, until the rotational decorrelation make the translation equivalent in each direction.}
\label{MSDScaling}
\end{figure}
The Brownian particle's diffusion constant in a homogeneous medium is inversely related to its hydrodynamic radius $R_H$. For a spherical particle, the hydrodynamic radius is the same as its radius. For the rod-like particles, the hydrodynamic radius scales as $R_H\sim\frac{L}{log(L/\sigma+b)}$ \cite{BookPoly}. Here, $L$ is the rod-length, $\sigma$ is the rod's width, and $b$ is the constant to account for the end effects. This expression for $R_H$ explains the rod-length scaling of the rod's CoM' diffusion constant and the relaxation time (discussed in the next section). Apart from this, the $MSD_{||}$ and $MSD_\perp$ would depend on rod length in a very different way. The parallel component should see the liquid from a view of the rod's diameter, i.e., $\sigma$. Thus, it should not depend on rod length at all. One can see in Fig.\ref{MSDScaling} (left-panel), the $MSD_{||}$ and $MSD_{||}/t$ collapse automatically for different rod-lengths in the 3dKA system at temperature $T=2.0$. On the other hand, the orthogonal component should see the liquid as a spherical particle of diameter $L$, implying the MSD should be scalable by $1/L$, as seen in Fig.1(right-panel).

\subsection{Diffusion constants of CoM of probe rods}
\begin{figure}[!b]	
\includegraphics[width=.950\textwidth]{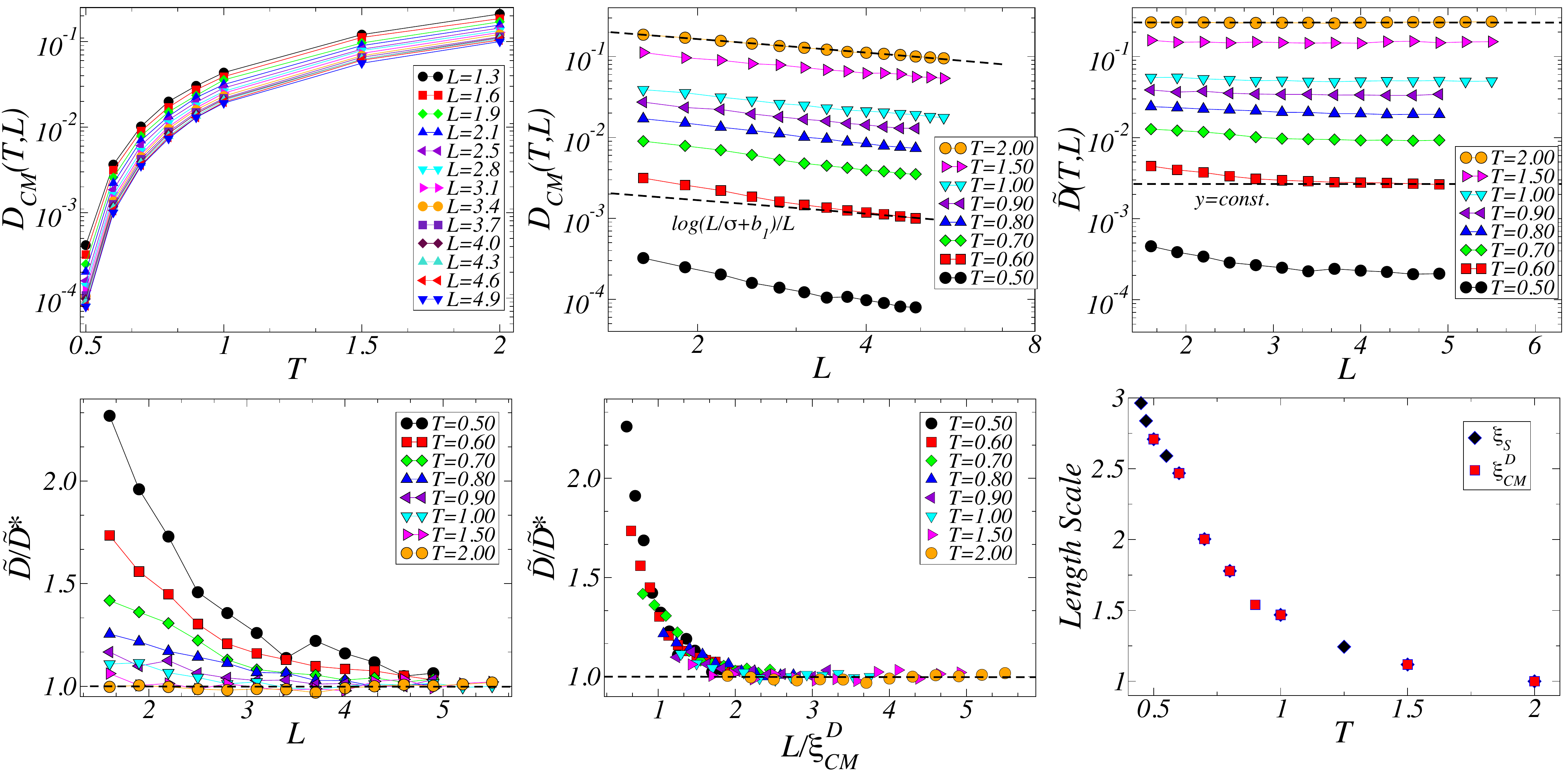}
\caption{Top-left panel: Temperature variation of diffusion constant 
of CoM of rod, $D_{CM}$ for rod of various lengths immersed in a liquid modelled 
by 3dKA model. In the top-center panel, the rod length dependence of $D_{CM}$ for 
different temperatures is plotted. The Brownian scaling of $D_{CM}\sim log(L/\sigma+b_1)/L$ fits the large temperature and large rod-length bits, perfectly. The fitted constant $b_1$ turns out to be, $b_1=1.49$ with $\sigma=1.00$ (width of the rod). This normal liquid scaling is removed by plotting $\tilde{D}(T,L)$ (defined in text) in top-right panel to observe the supercooling effect. With increasing supercooling the larger and larger rods couldn't survive the emerging structural correlations. In the bottom-left panel the $\tilde{D}/\tilde{D}^*$ is plotted, where $\tilde{D}^*$ is the large rod-length asymptotic value of $\tilde{D}$. Here the effect of supercooling can be clearly seen, its interpretation is given in text.   $\tilde{D}/\tilde{D}^*$ is collapsed (bottom-center) to obtained the underlying growing length scale, $\xi_{CM}^D$ (bottom-right). The temperature dependence of $\xi_{CM}^D$ is also compared with the static length scale of the system $\xi_S$, obtained as PTS length scale.}
\label{TransDiff3dKA}
\end{figure}
\begin{figure}[!h]	
\includegraphics[width=.950\textwidth]{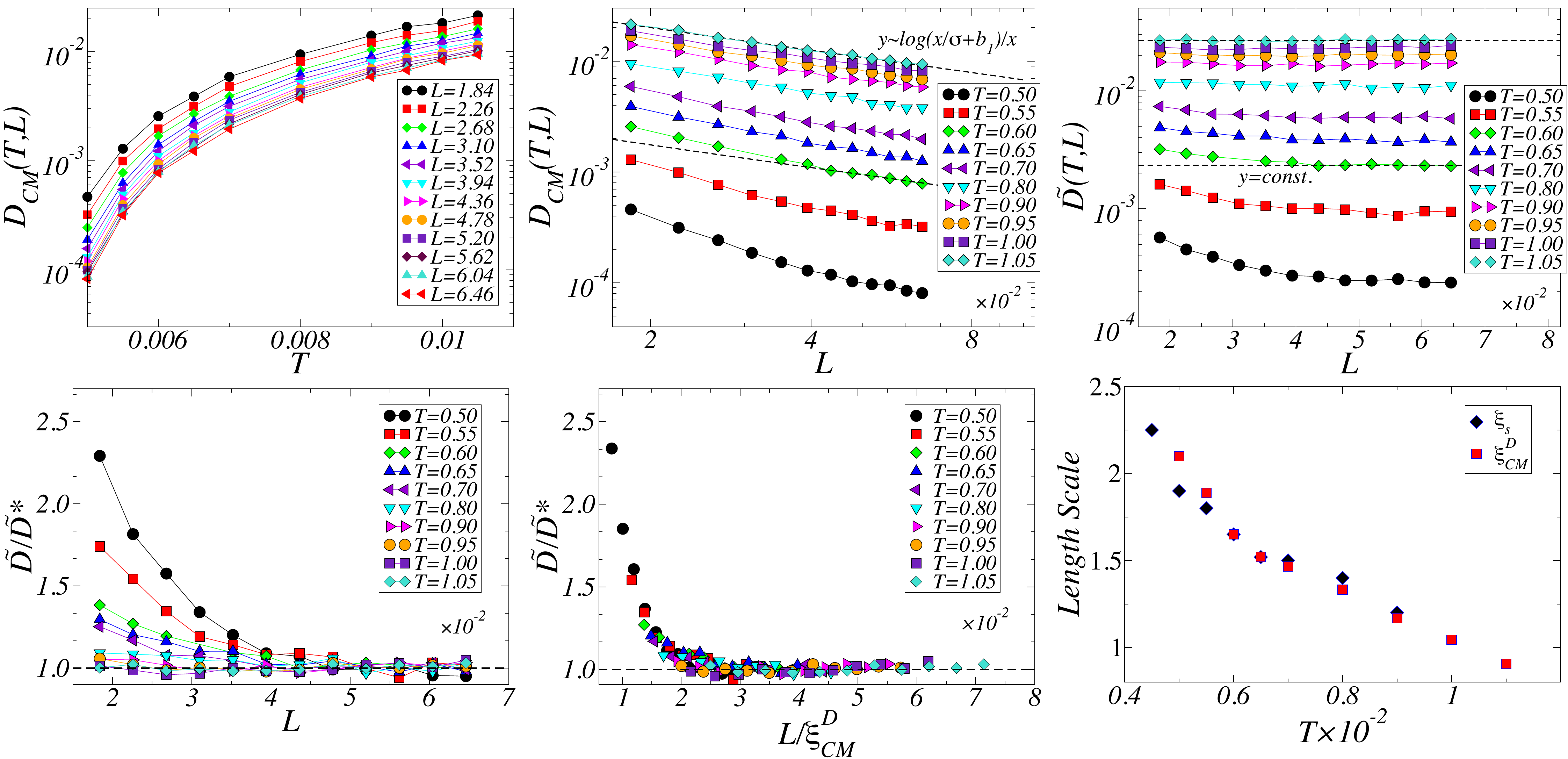}
\caption{Similar quantities are plotted as in Fig.\ref{TransDiff3dKA} for the 3dHP model to reach similar conclusions. The fitting constant $b_1$ turns out to be $b_1=2.05$}.
\label{TransDiff3dHP}
\end{figure}

Translation diffusion constant ($D_{CM}$) for the CoM of the rod can be obtained 
from MSD using an expression $D_{CM}(T,L)=MSD(t\rightarrow\infty,T,L)/6t$. 
Temperature variation of $D_{CM}$ for rods of various lengths in 3dKA (and 3dHP) liquid 
and its rod length variation for rods in different temperature liquid is plotted in the top-left 
and top-center panel of Fig.\ref{TransDiff3dKA} (and Fig.\ref{TransDiff3dHP}), respectively. 
As discussed in the previous section and in the main text, the diffusion constant for a Brownian 
rod should follow the $ln(L/\sigma+b_1)/L$ dependence with $b_1$ being a constant. 
Fig.\ref{TransDiff3dKA} \& \ref{TransDiff3dHP} (top-center panel) clearly validate the scaling 
for rods in high temperature 3dKA \& 3dHP liquid. Also, In the top-right panel of the respective 
figures, we removed this asymptotic rod length dependence by plotting 
$\tilde{D}(T,L)=D_{CM}L/ln(L/\sigma+b_1)$ w.r.t. rod length. In our models, $\sigma=1.0$, 
and the value of $b_1$ is obtained by fitting the high-temperature data. One can clearly see 
the effect of supercooling in play, the diffusion-constant, $\tilde{D}$ starts to deviate from 
the constant value at larger and larger rod length with increasing supercooling. This indicates 
the growth of some underlying length-scale. The deviation here implies the faster diffusion 
of the smaller rods, while the larger rods still feel the liquid as a homogeneous medium. As 
argued in the main text and in Ref.~\cite{Anoop}, this increase of deviation in $\tilde{D}$ from
its constant high temperature and large rod length value with decreasing probe-length is linked 
to the parent liquid's correlated relaxation processes, the linear-length of which gives the static 
length scale of the system. In Fig.\ref{TransDiff3dKA} \& \ref{TransDiff3dHP} (bottom-left panel) 
we scaled $\tilde{D}$ by its large rod-length value at each temperature, i.e., 
$\tilde{D}^*(T)=\tilde{D}(T,L\rightarrow\infty)$. This data is then collapsed to obtain the 
underlying length scale ($\xi_{CM}^D$) using the following scaling ansatz.
\begin{equation}
\frac{\tilde{D}(T,L)}{\tilde{D}^*(T)}=\mathscr{F}\left(\frac{L}{\xi_{CM}^D}\right)
\label{scalingD}
\end{equation}
The collapse and the comparison of obtained length scale with the static length scale obtained 
via PTS correlation method is shown in Fig.\ref{TransDiff3dKA} \& \ref{TransDiff3dHP} 
(bottom-center, right panel) for 3dKA and 3dHP model system, respectively. The quality of collapse 
and the astonishing similarity in the temperature dependence of obtained length-scale with a 
static length scale of liquid prove the procedure's robustness. The whole procedure is summed-up 
in Figs.\ref{TransDiff3dKA} \& \ref{TransDiff3dHP} for 3dKA and 3dHP models respectively.


\subsection{Translational relaxation of CoM of rod}
\begin{figure}[!h]	
\centering
\includegraphics[width=.95\textwidth]{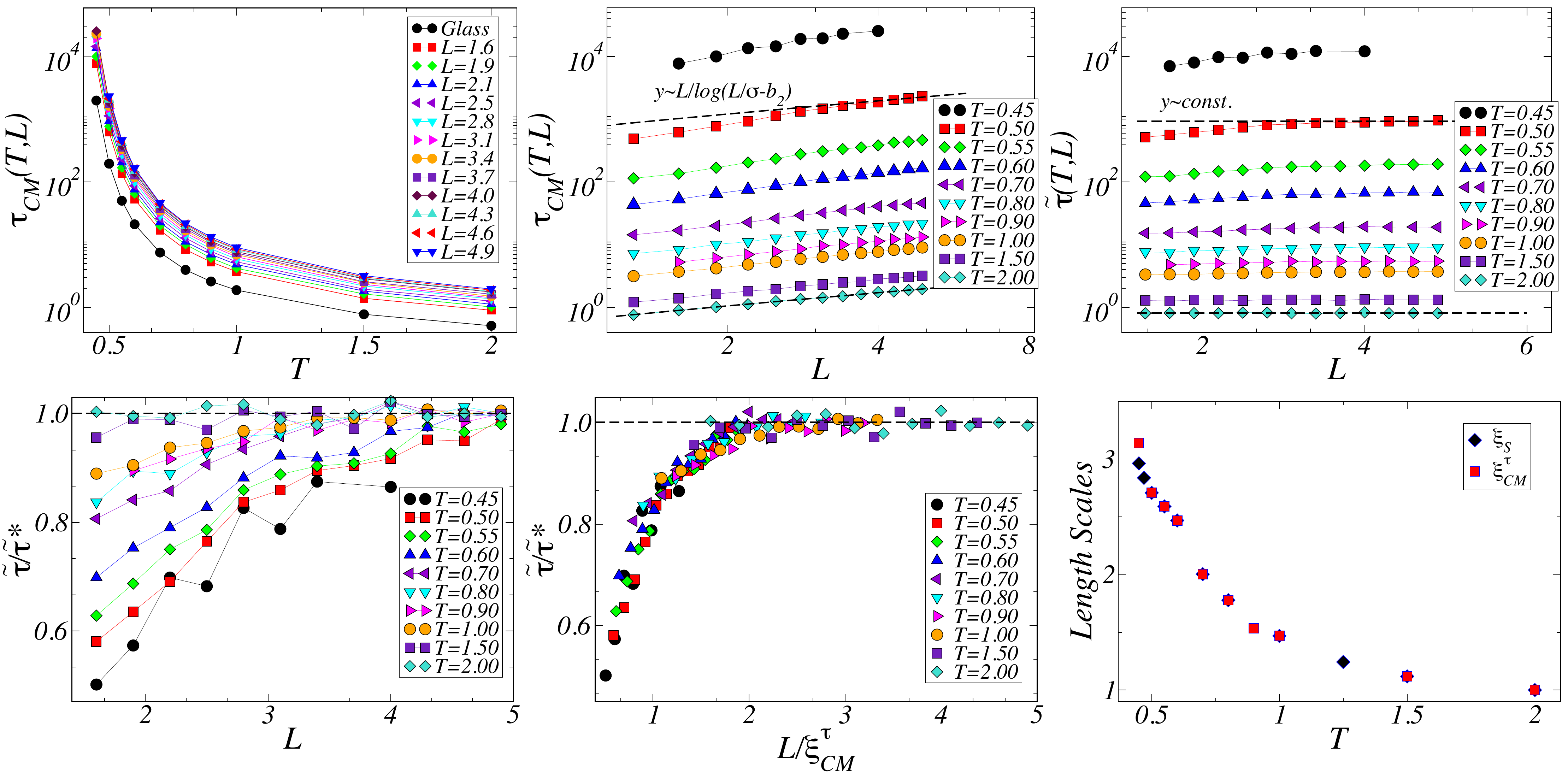}
\caption{ The temperature dependence and the rod-length dependence of translational relaxation time, $\tau_{CM}(T,L)$ for rod of different lengths and rods in different temperature 3dKA model are plotted in top-left and top-right panels, respectively. The Brownian rod length dependence of $\tau_{CM}\sim L/log(L/\sigma+b_2)$, is well obeyed for the high temperature liquid and for larger rod-length. The fitting parameter $b_2$, turns out to be $b_2=2.67$ with $\sigma=1.0$ being the width of the rod. $\tilde{\tau}(T,L)$ is then plotted in the top-right panel with this scaling removed, i.e, $\tilde{\tau}=\tau_{CM}\times \frac{log(L/\sigma+b_2)}{L}$. Any deviation from $\tilde{\tau}(T,L)=f(T)$ is an effect of supercooling. In bottom-left panel, $\tilde{\tau}/\tilde{\tau}^*$ is plotted, where $\tilde{\tau}^*(T)$ is the large rod-length asymptote value of $\tilde{\tau}(T,L)$. This quantity is then collapsed in the bottom-middle panel to obtain the underlying growing structural order ($\xi_{CM}^D$). Temperature dependence and its comparison with the system's static length scale obtained as PTS length scale is presented in the bottom right panel.
}
\label{TransRelax3dKA}
\end{figure}

\begin{figure}[!ht]	
\centering
\includegraphics[width=.95\textwidth]{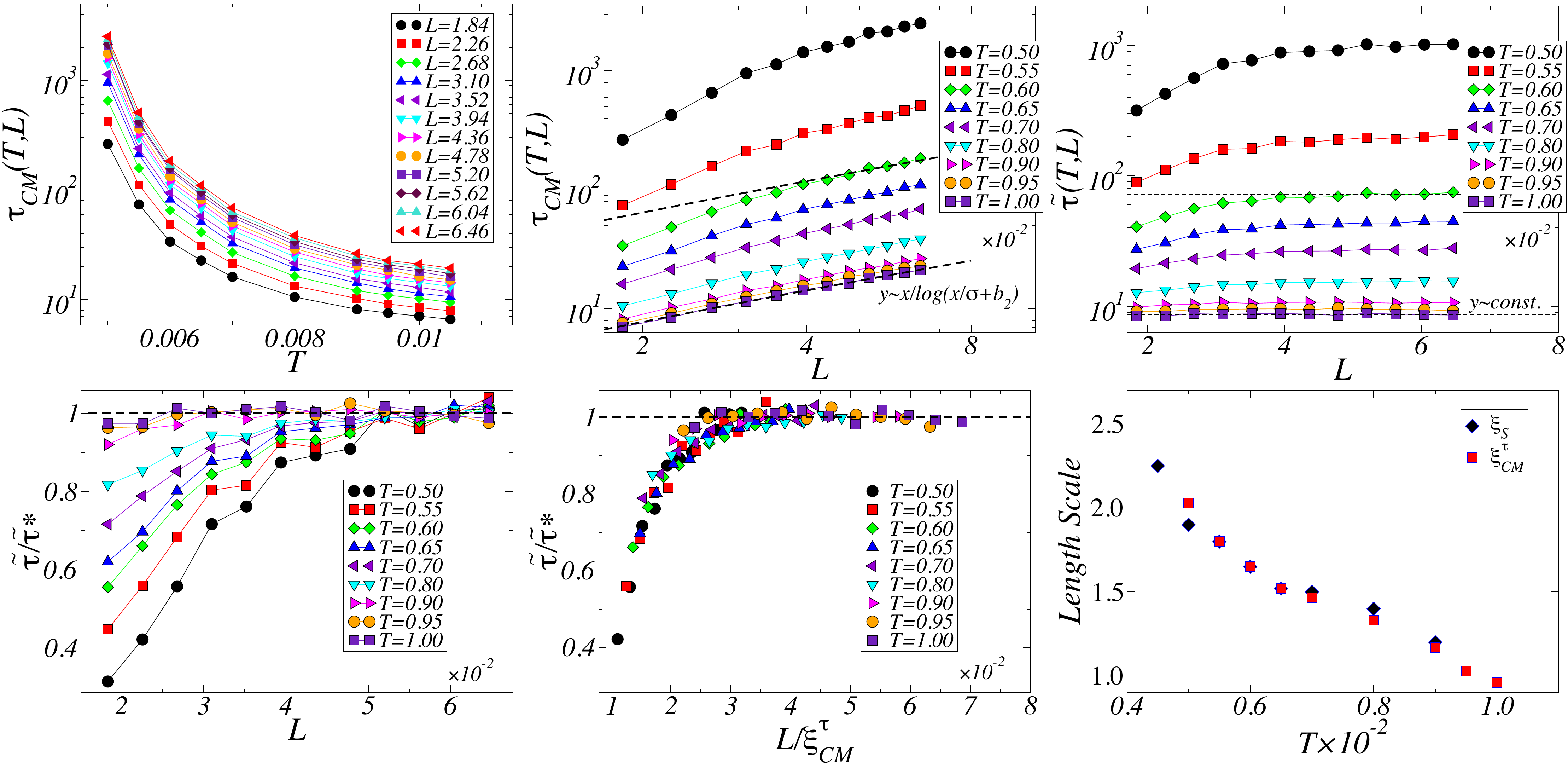}
\caption{Similar quantities are plotted as of Fig.\ref{TransRelax3dKA} for 3dHP model to reach similar conclusions. The fitting constant $b_2$ turns out to be $b_2=6.10$. }
\label{TransRelax3dHP}
\end{figure}

The translational relaxation time of a rod is defined as the time when overlap function, $Q(t)$ for CoM of a rod (Eq:\ref{overlap}) reduces to $1/e$, i.e., $\langle Q(t=\tau_{CM})\rangle=1/e$. In Fig.\ref{Qt}, $Q(t)$ is plotted for rod of length $L = 2.5$ \& $L=3.1$ immersed in the 3dKA \& 3dHP liquid at different temperatures. Temperature dependence of $\tau_{CM}$ for various sized rods and its rod length dependence in the 3dKA (3dHP) liquid at different temperatures can be found in the left and center panel of Fig.\ref{TransRelax3dKA} (Fig.\ref{TransRelax3dHP}) respectively.  The rod length scaling of $\tau_{CM}(T,L)$ should be inverse to that of $D_{CM}$, thus following the similar route as to extract the static length scale form diffusion constant, we can obtain the static length scale. The quantity of interest here would be, $\tilde{\tau}(T,L)=\tau_{CM}ln(L/\sigma+b_2)/L$, which would be scaled by large rod-length value at each temperature i.e., $\tilde{\tau}^*(T)=\tilde{\tau}(T,L\rightarrow\infty)$. Then on using the following scaling ansatz the relevant length scale ($\xi_{CM}^{\tau}$) would be extracted. 
\begin{equation}
\frac{\tilde{\tau}(T,L)}{\tilde{\tau}^*(T)}=\mathscr{F}\left(\frac{L}{\xi_{CM}^\tau}\right)
\label{scalingTau}
\end{equation}
This procedure is outlined in Fig.\ref{TransRelax3dKA} \& \ref{TransRelax3dHP} for 3dKA and 3dHP model system.


\section{Stokes-Einstein Relations}
\begin{figure}[!ht]	
\centering
\includegraphics[width=1.0\textwidth]{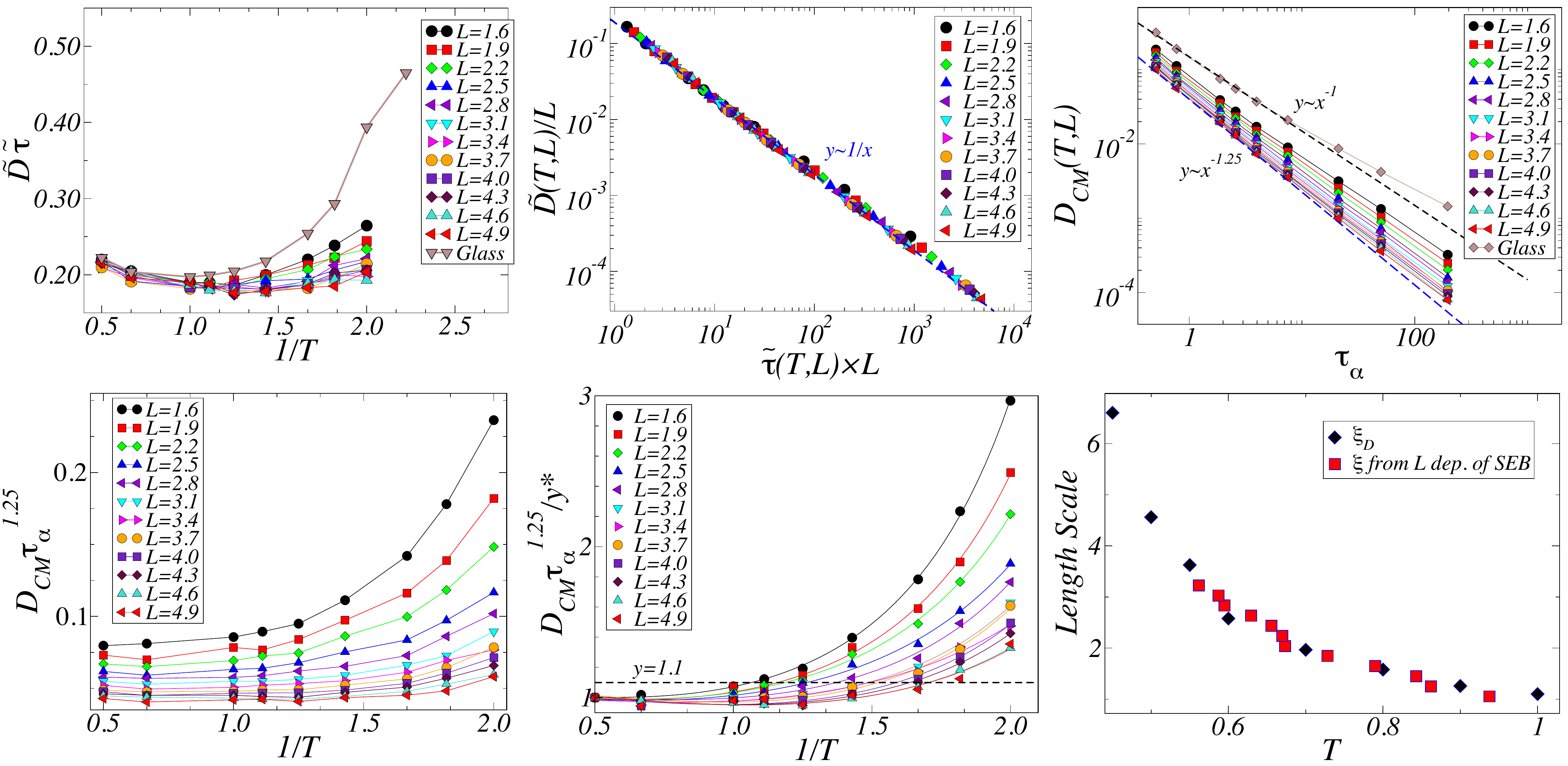}
\caption{Top panel: In the left, scaled SE parameter $\tilde{D}\tilde{\tau}$ is plotted with increasing supercooling for different rod lengths 
along with the similar SE parameter plot for parent liquid. The non-violation 
of this SE parameter is displayed best in the center panel with cross plot of 
$\tilde{D}/L$ and $\tilde{\tau}\times L$. In the right panel similar cross plot between 
$D_{CM}$ and $\tau_\alpha$ suggests the power law of $D_{CM}
\sim\tau_\alpha^{-1.25}$ emerging for large rods and high temperature. 
Bottom panel: The fractional SE parameter $D_{CM}\tau_\alpha^{1.25}$ is plotted in 
the left panel for different rods with respect to inverse temperature. The SEB 
for this parameter happens at lower and lower temperature for larger rods. The 
SEB temperature for a particular rod length is chosen to be the temperature 
when $D_{CM}\tau_{\alpha}^{1.25}/y^*=1.1$, while this quantity is equal to 
unity for high temperature. This is outlined in center panel of the figure 
with solid lines are fit to the function mentioned in text. The obtained rod 
length dependence for SEB temperature is compared with dynamic length scale of 
the liquid in right panel.       
}
\label{SETrans3dKA}
\end{figure}

\begin{figure}[!ht]	
\centering
\includegraphics[width=1.0\textwidth]{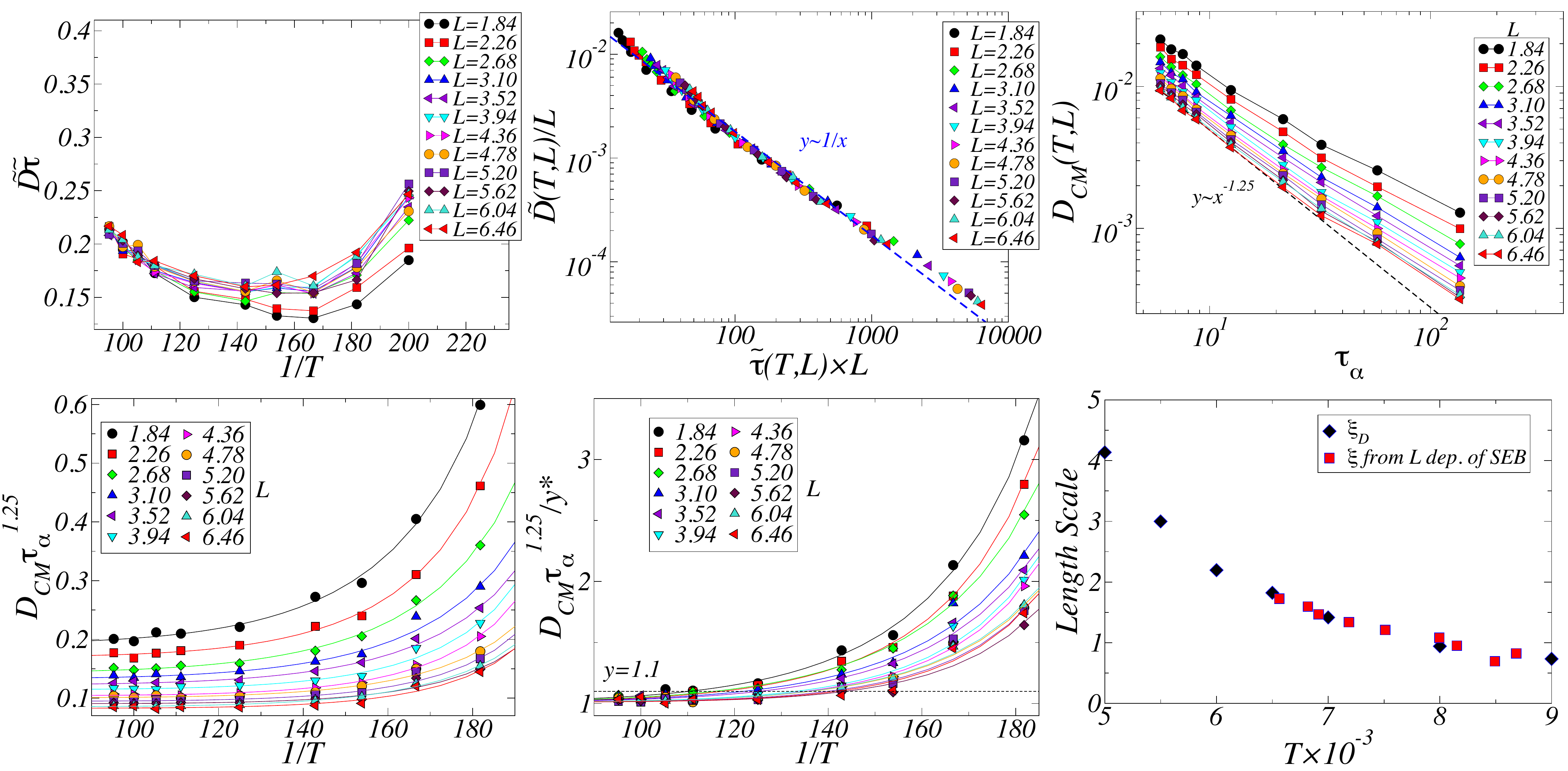}
\caption{Similar quantities are plotted as in Fig.\ref{SETrans3dKA} for 3dHP model system.
}
\label{SETrans3dHP}
\end{figure}
The Stokes-Einstein (SE) relation is a blend of Einstein's work on diffusion for 
spherical particles and Stokes's study on drag felt by a spherical body through the 
viscous medium, and it reads as,
\begin{equation}
D=\frac{k_B T}{c\pi R\eta}
\end{equation}
where $D$ and $R$ respectively are the diffusion constant and the hydrodynamic 
radius of the particle, $\eta$ and $T$ are the shear viscosity, and the temperature 
of liquid respectively, and $c$ is a constant which depends on the boundary 
condition at the surface of the particle ($c=6$ for stick \& $c=4$ for slip) 
\cite{LandauFD}. This expression is also valid for the liquid particle's self-diffusion, 
and Stokes-Einstein-Breakdown (SEB) is extensively studied as a possible effect 
of the presence of dynamic heterogeneity in the system. Since the distribution of 
local viscosity (structural relaxation time) gets broader due to dynamic 
heterogeneity, the diffusion constant's growth becomes faster than the predicted 
value from the SE relation. In this mathematical equation, viscosity is often replaced by 
relaxation time ($\eta\sim\tau$) or temperature times relaxation time ($\eta\sim 
\tau T$) as they have very similar temperature dependence at least deep in the 
supercooled temperature regime where viscosity grows very rapidly with small 
change in temperature. The first one grounds its correctness in 
Maxwell relation, i.e., $\eta=G_\infty\tau$ where $G_{\infty}$ is the instantaneous 
shear modulus. So the interchangeability of viscosity and relaxation time relies on 
the assumption of $G_{\infty}$ being temperature independent, but it can have 
some temperature dependence after all. The second relation comes from the 
Gaussian solution of the diffusion equation, which links the diffusion constant with 
relaxation time as $D_tk^2=1/\tau_t$, where $k$ is the wave vector associated 
with the self-intermediate scattering function. Both of these proxies are well 
studied in the literature, and one can find their comparison in \cite{Shi2013}, but 
since we would be using SE relation for rods (more Brownian-like behaviour), the 
choice $\eta\sim\tau T$ seems more physical. Thus the Stokes-Einstein relation 
reads out,
\begin{equation}
D\tau=\frac{1}{c\pi R}=f(R).
\end{equation}  

$f(R)$ can be considered as a measure of SEB. For rods in liquid, we can have 
$D_{CM} \tau_{CM}$ and $D_{CM} \tau_\alpha$ as SEB parameters where, as 
defined earlier, $D_{CM}$ \& $\tau_{CM}$ is the diffusion constant and relaxation 
time of CoM of the rod, and $\tau_\alpha$ is the relaxation time of liquid. Fig.
\ref{SETrans3dKA} \& Fig.\ref{SETrans3dHP} top-left panel shows the temperature 
variation of $\tilde{D} \tilde{\tau}$ for different sized rods. As expected from the 
previous scaling analysis of $D_{CM}$ and $\tau_{CM}$, that product of $\tilde{D}
\tilde{\tau}$ would be a scaling function of the static length scale (using Eq.
\ref{scalingD} \& Eq.\ref{scalingTau}). It turns out that scaling functions that govern 
the rod length dependence of $\tilde{D}$ and $\tilde{\tau}$ are inverse of each other. 
Thus $\tilde{D}	\tilde{\tau}$ will be constant. The smallness of the extent of 
breakdown can be seen in Fig.\ref{SETrans3dKA} \& Fig.\ref{SETrans3dHP}'s 
top-center panel (3dKA \& 3dHP), where $\tilde{D}\sim\tilde{\tau}$ for a given rod 
length is confirmed in their cross plot. A similar cross plot between $D_{CM}$ and $
\tau_\alpha$ is also included in the top-right panel of the same figure, which predicts 
a different power law of $D_{CM}\sim\tau_\alpha^{-1.25}$ for high temperature and 
large rods. As pointed out in the main text, the temperature of departure from this 
power-law or the SEB temperature for a different rod length behaves like the 
dynamic length scale of the system \cite{SastrySE}.  This procedure, in complete 
depth, is illustrated in Fig.\ref{SETrans3dKA} \& Fig.\ref{SETrans3dHP} bottom 
panels for the 3dKA and 3dHP models, respectively. The SEB temperature for 
different rods is the value of $1/T$ at which $y/y^*=1.1$. Here $y(T,L)=D_{CM}\tau_
\alpha^{1.25}$ and $y^*(L)=y(2.0,L)$. This procedure is very similar to the one 
adopted in Ref.\cite{SastrySE}. The solid lines in these figure's bottom-center panel 
are the fit to function $ln(y/y*)=[x+c_0(x/x_0)^n+c_1(x/x_0)^{2n}]$ where $x=1/T$.

\section{Rotational Quantities}
\subsection{Rotational correlation function}
The $l^{th}$ order rotational correlation function in three dimensions is defined as, 
\begin{equation}
g_l(t)=\langle P_l(cos[\theta(t)])P_l(cos[\theta(0)])\rangle
\label{Rotcorr}
\end{equation}
Where $P_l$ is the $l^{th}$ order Legendre Polynomial. In the single-molecule linear-dichroism experiments one get the major contribution to the auto-correlation function from $g_2(t)$, which in-turn is linked to the rotational diffusion constant as $g_{l}(t)=exp(-D_rl(l+1)t)$. Also the non-exponential decay of these correlation function is the identification of dynamical heterogeneity in the system.
\subsection{Rotational Non-Normal Parameter}
The rotation non-normal parameter (NNP),  $\alpha_{3D}^{rot}(t)$ is a function of fourth and second order moments for rotational motion combined in way to give zero for homogeneous diffusion. In other words in heterogeneous system, NNP would be a measure of fluctuation in the rotational diffusion constant at that time. For three dimensions following is the NNP \cite{Jain2017}. Validity of the same can be checked by solving the rotational diffusion equation and finding the respective moments. 
\begin{equation}
\alpha_{3D}(t) = \frac{1}{2}\frac{\left\langle|\hat{u}(t)-\hat{u}_0|^4\right
\rangle}{\left(\left\langle|\hat{u}(t)-\hat{u}_0|
^2\right\rangle\right)^2}
+\frac{1}{6}\left\langle|\hat{u}(t)-\hat{u}_0|^2\right\rangle\\ -1
\end{equation}

\section{Stokes-Einstein-Debye Relation}  
\begin{figure*}[!h]	 
\centering
\includegraphics[width=0.950\textwidth]{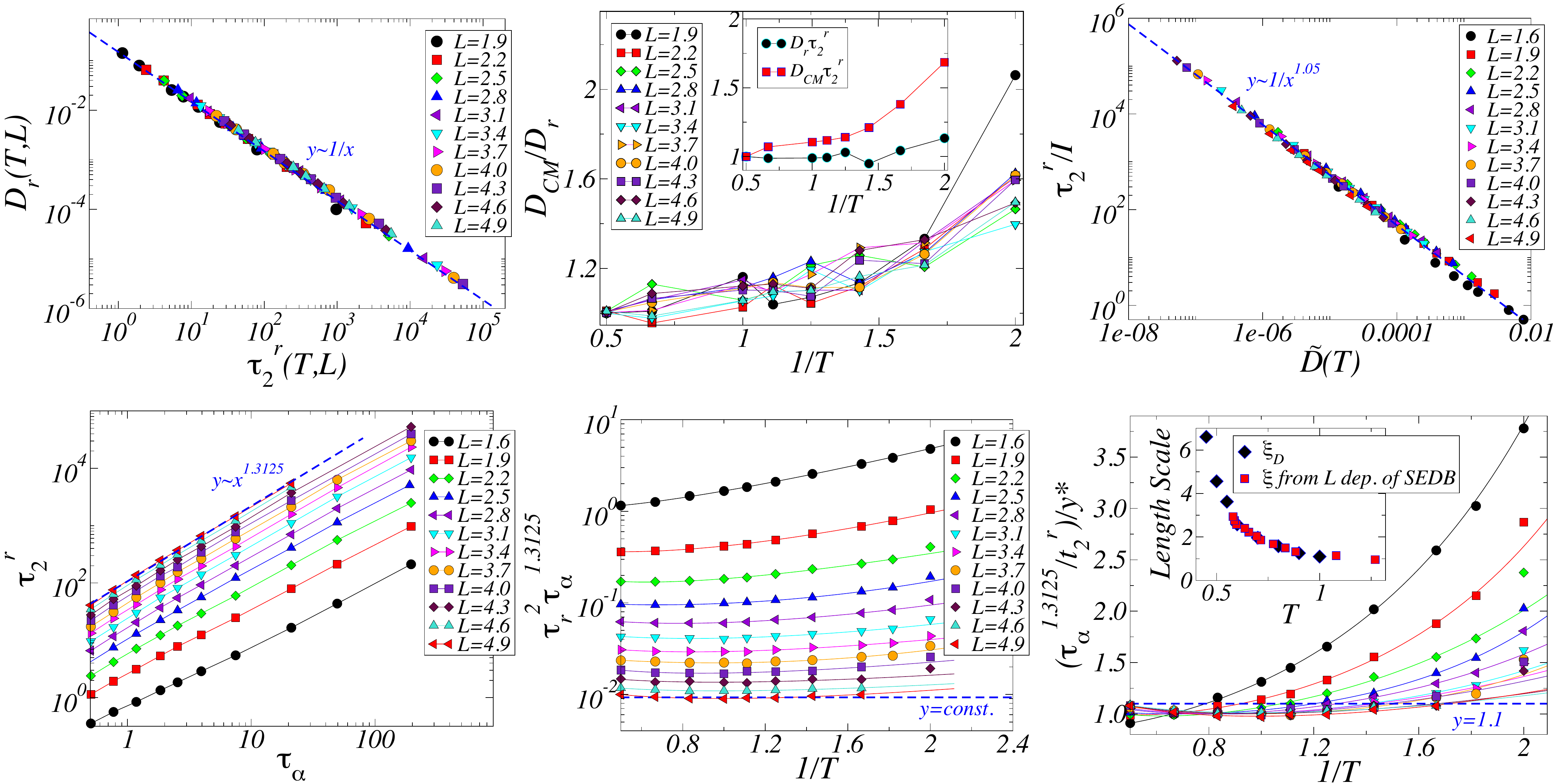}
\caption{In top-left panel the Debye relation for rotational diffusion constant and rotational relaxation time, $D_{r}(T,L)\sim \frac{1}{\tau_2^r(T,L)}$ is tested in their cross plot for rods in 3dKA model system.  In top-center panel, the ratio $D_{CM}/D_r$ is plotted w.r.t. inverse 
temperature. According to SED relation this ratio should remain constant with 
temperature,  
thus its increase with decreasing temperature would lead to a 
conclusion of decoupling of translational-rotational dynamics of the probe rods.
This supercooling effect of enhancement in the translational motion as compared to rotational motion is also inferred from $D_r\tau_2^r$ and 
$D_{CM}\tau_2^r$ plots for the rod of length $L=4.5$ in the inset. 
The cross plot of $\tau_2^r/I$ and $\tilde{D}(T,L)$ in the top-right panel predicts the power-law of $\tau_2^r\sim D_{CM}^{-1.05}$, which is again consistent with the enhancement of translation with supercooling. This direct link also enable us to perform the same SE-breakdown analysis on SED-breakdown to extract the dynamic length scale. The same for 3dKA system is highlighted in the bottom panels.  
}
\label{SED3dKA}
\end{figure*}
\begin{figure*}[!h]	 
\centering
\includegraphics[width=0.950\textwidth]{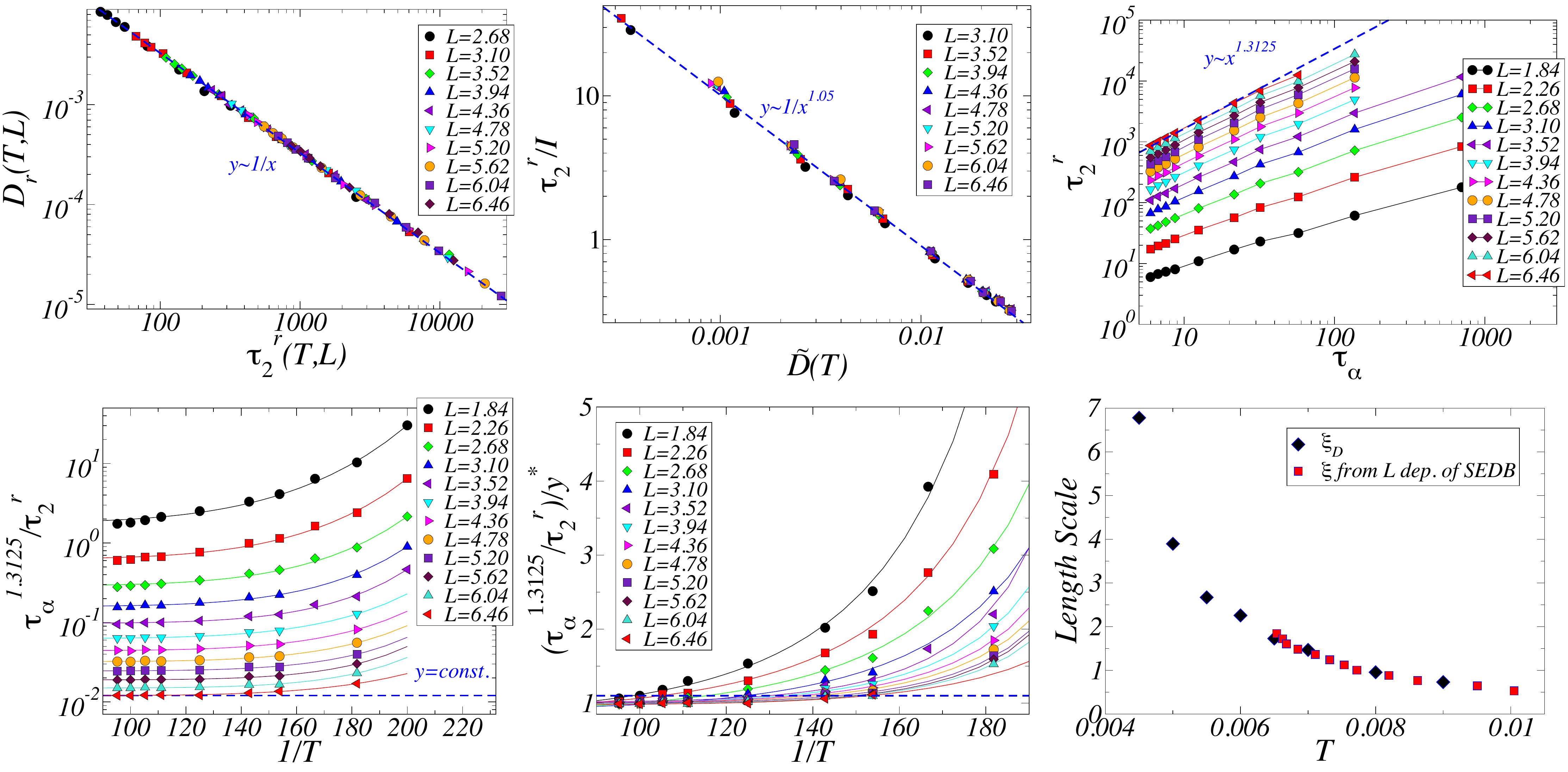}
\caption{Similar quantities are plotted as in Fig.\ref{SED3dKA} for 3dHP model system.}
\label{SED3dHP}
\end{figure*}

For rotational dynamics we have the Debye relation linking the rotational diffusion 
constant ($D_r(T,L)$) with $l^{th}$ order rotational correlation time $\tau^r_l$  which is 
defined as the time when $l^{th}$ order rotational correlation function $g_l(t,T)$ 
(Eq:\ref{Rotcorr}) reduces to $1/e$. The Debye relation reads, $D_r=1/\tau^r_l l (l+1)
$, implying the factor $D_r\tau_l^r$ is independent of temperature. In top-left 
panel of Fig.\ref{SED3dKA} \& Fig.\ref{SED3dHP} this has been checked in the cross plot of 
$D_r$ and $\tau_2^r$ for various rod lengths in 3dKA \& 3dHP model respectively. The $D_r
\sim 1/\tau_2^r$ suggests that the Debye relation has survived the supercooled 
environment. Now if we combine the Debye relation with rotational SE relation i.e. 
$D_r=k_BT/8\pi\eta R^3$ then we would have the following Stokes-Einstein-Debye (SED) 
relation, 
\begin{equation}
\frac{1}{\tau_r^l l (l+1)}=\frac{k_BT}{8\pi\eta R^3}
\end{equation}
and with the replacement of $\eta$ with relaxation times, we have $\frac{\tau_{l}^r}{\tau_{CM}}$ 
and $\frac{\tau^{l}_r}{\tau_{\alpha}}$ as the SED parameters. Now as discussed in the main text, the translational motion gets decoupled from the rotational motion with supercooling. This phenomenon can been seen in Fig.\ref{SED3dKA} (top-center) where ratio the $D_{CM}/D_{r}$ grows with supercooling, regardless of the rod size. Also, the same can be verified from the cross plots of $\tau_2^r$ and $D_{CM}$ (top-right panel of Fig.\ref{SED3dKA} \& Fig.\ref{SED3dHP}). This implies that the self-SED relations are violated with increasing supercooling regardless of the probe size. From the observed power laws namely, $\tau_2^r\sim D_{CM}^{-1.05}$ and $D_{CM}\sim \tau_\alpha^{-1.25}$, one expects $\tau_2^r\sim \tau_\alpha^{1.3125}$ along with similar information of dynamic length scale contained in the rod-length dependence of the breakdown-temperature for this relation. The analysis for SED-Breakdown is presented in Fig.\ref{SED3dKA}  \& Fig.\ref{SED3dHP} for 3dKA and 3dHP model systems, along with the comparison of length scale obtained. The rotational relaxation time $\tau_2^r(L,T)$ is easy to obtain experimentally  than the translational quantities hence gives an advantage to the analysis.
\section{Existing experimental studies}
\begin{figure}[!h]
\includegraphics[width=0.95\textwidth]{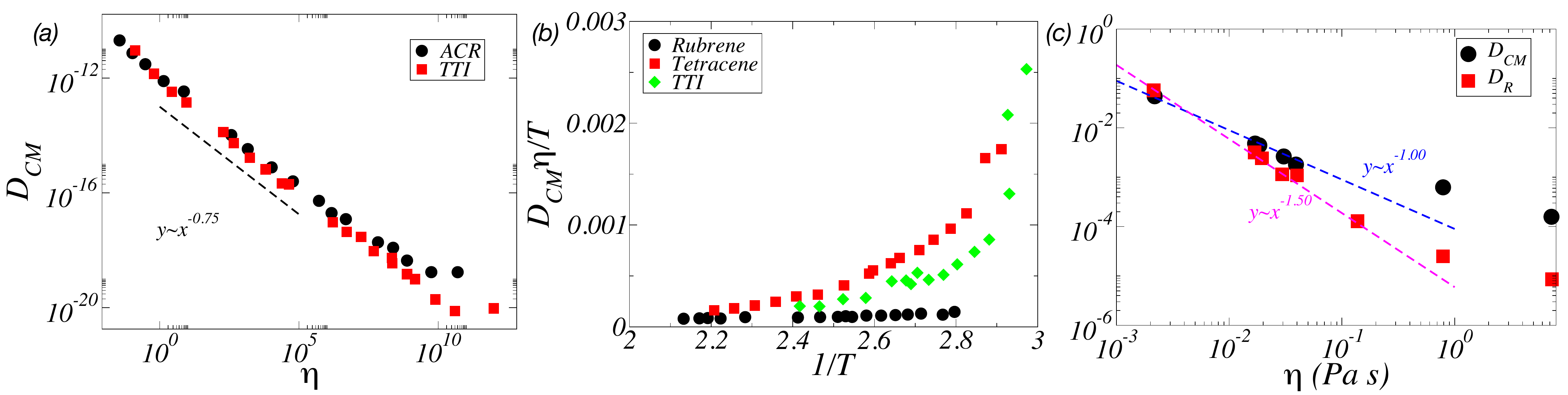}
\caption{(a) The $D_{CM}$ of probes is plotted against the viscosity $\eta$ of supercooled \textit{OTP} \cite{ExptOTPSalol1994}. One can clearly see the emerging power law and the deviation from it 
for different probes. The result is very similar to what we were expecting. (b) The SE parameter 
$D_{CM}\eta/T$ is plotted for three different probes in supercooled \textit{TNB} \cite{Blackburn}. Again 
the results are in agreement of our simulation results. Note that hydrodynamic radius of \textit{Rubrene} is 
larger than \textit{TTI } while hydrodynamic radius of \textit{Tetracene} is smaller than \textit{TTI}. The breakdown 
temperature is then expected to be smallest the largest probe size {\it i.e. Rubrene}. We also highlight
that this current experimental dataset is not enough to get a good estimate of the growing heterogeneity 
length scale but it certainly proves its existence. (c) The $D_{CM}$ 
and $D_R$ of tracer clusters are plotted against $\eta$ of supercooled polymer fluid to highlight 
the different power-laws and deviation from it with increasing supercooling \cite{PolyWeeks}.   }
\label{Expt}
\end{figure}
As mentioned in the main text, there are many single molecule experiments in literature which concentrate 
on the translation along with rotation of the probe molecule to quantify dynamic heterogeneity in supercooled 
liquids. The experiments reported mostly the variation of translational ($D_{CM}$) and rotational ($D_{R}$) 
diffusion constant  with shear viscosity $\eta(T)$. Since $\eta(T)$ for supercooled liquids is known to show
diverging behaviour upon approaching the glass transition temperature, $T_g$, so the question usually 
addressed was, is $D_{CM}$ \& $D_{R}$ also follows the $T/\eta$ or they get decoupled? 
In Ref~\cite{ExptOTPSalol1994} the same was checked for various probes in \textit{othoterphenyl (OTP)} and 
\textit{salol}. It was found that $D_{CM}$ have weaker temperature dependence of $D_{CM}\sim \eta^{-0.75}$ 
for probes than the rotational diffusion constant, $D_{R}\sim \eta^{-1}$ for self \textit{OTP} particles. This result 
is shown in Fig.\ref{Expt} (a) for two different probes,\textit{ Aberchrome 540 (ACR) and 
2,2-bis(4,4-dimethylthiotan-3-one (TTI}) in supercooled \textit{OTP}. One can very clearly see the effect of 
probe size in the deviation from this power-law, as proposed in our work. 
This experiment uses H-NMR stimulated echo method in static magnetic gradients. In another work 
by Blackburn et al.\cite{Blackburn} the SE quantity $D_{CM}\eta/T$ was calculated for various probes 
in supercooled \textit{1,3,5-Tris(naphthyl)benzene (TNB)}. In Fig.\ref{Expt}(b), one can clearly see the effect of 
probe size here as well in the SEB temperature. The probes used were \textit{rubrene}, t\textit{etracene} and \textit{TTI}.  
Note that \textit{rubrene} has larger hydrodynamic radius than \textit{TTI} while \textit{tetracence} has smaller hydrodynamic 
radius than \textit{TTI}. Thus it is expected that \textit{tetracence} will have SEB temperature that is much higher than
either \textit{TTI} or \textit{rubrene}. This is in complete agreement with what is observed in our simulations. Thus it is 
clear that systematic experimental measurements on various probes with changing linear size will 
definitely lead to extraction of dynamic heterogeneity length scale in these molecular glass-forming 
liquids. We also want to highlight that this extracted data from existing experimental results are not
good enough to extract the underlying length scale but it clearly demonstrate that our proposal is
very robust and experimentally feasible. Lastly the deviation form the power laws $D_{CM}\sim \eta^{-1.0}$ 
and $D_{R}\sim \eta^{-1.50}$ with increasing supercooling can also be seen for tracer clusters in colloidal 
supercooled fluids in Fig.\ref{Expt} (c). This data is adapted from Ref.\cite{PolyWeeks}.

In the main text we illustrated a table containing various probes already been used in various supercooled liquids. The full description table with full chemical names is provided in the following table.
\begin{center}
\begin{tabular}{ | p{3.5cm} | p{12.5cm} | }
\hline
Medium & Probe \\ \hline
	\multirow{5}{*}{\textit{Glycerol}}	& (a) \textit{N,N$'$-bis(2,5-\textit{tert}-butylphenyl)-3,4,9,10-perylenedicarboximide (tbPDI)}
										\cite{Mackowiak2011,Zondervan2007} \\
										& (b) \textit{N,N$'$-bis-(2,6-dimethylphenyl)-3,4,9,10-perylenedicarboximide (dpPDI)}
										\cite{Mackowiak2011} \\
										& (c) \textit{N,N$'$-bis[3-(dimethylamino)propyl]-3,4,9,10-perylenedicarboximide (dapPDI)}
										 \cite{Mackowiak2011} \\
										& (d) \textit{rubrene} molecule \cite{Mackowiak2009} \\
										& (e) Gold nano-rods \cite{Yuan2013} \\ \hline
	\multirow{9}{*}{\textit{ortho-terphenyl (OTP)}} &	(a) \textit{rubrene} \cite{Cicerone1995,Cicerone1996} \\
										& (b) \textit{9,10-bis(phenylethynyl)anthracene (BPEA)} \cite{Cicerone1996} \\
										& (c) \textit{tetracene} \cite{Cicerone1993, Cicerone1995,Cicerone1996} \\
										& (d) \textit{anthracene} \cite{Cicerone1993,Cicerone1996} \\
										& (e) \textit{dpPDI}\\
										& (f) \textit{tbPDI}\\ 
										& (g) \textit{N,N$'$-bis(triethylglycol)-3,4,9,10-perylenedicarboximide (egPDI)}\\
										& (h) \textit{N,N$'$-dipentyl3,4,9,10-perylenedicarboximide (pPDI)} \cite{Leone2013}\\
										& (i) \textit{BODIPY268} dye \cite{Paeng2015}\\ \hline								
\end{tabular}
\end{center}
\section{Model Systems \& simulation details}
\noindent{\bf 3dKA Model:}
The 3dKA model is a generic model to mimics the molecular liquids ($Ni_{80}P_{20}$ glass-forming liquids).
It comprise of 80:20 binary mixture of A and B type Lennard-Jones particles.
The interaction potential is given by
\begin{equation}
V_{\alpha\beta}(\textbf{r})=4\epsilon_{\alpha\beta}
\left[\left(\frac{\sigma_{\alpha\beta}}{r}\right)^{12}-\left(\frac{\sigma
_{\alpha\beta}}{r}\right)^6\right]
\end{equation}
where $\alpha$ and $\beta$ varies in A, B and the interaction strengths and radii 
are $\epsilon_{AA}=1.0$, $\epsilon_{AB}=1.5$, $\epsilon_{BB}=0.5$; $
\sigma_{AA}=1.0$, $\sigma_{BB}=0.88$ and $\sigma_{AB}=0.8$. The interaction 
is truncated at $r=2.5\sigma_{\alpha\beta}$ and is made smooth by adding upto 
$4^{th}$ order terms. The simulations are done in temperature range $\
{T=0.45,1.00}$ while the pressure is chosen to maintain the average number density 
$\rho=1.2$ with $N=5000$ particles.\\
\begin{figure}[!hpt]
	\includegraphics[width=0.8\textwidth]{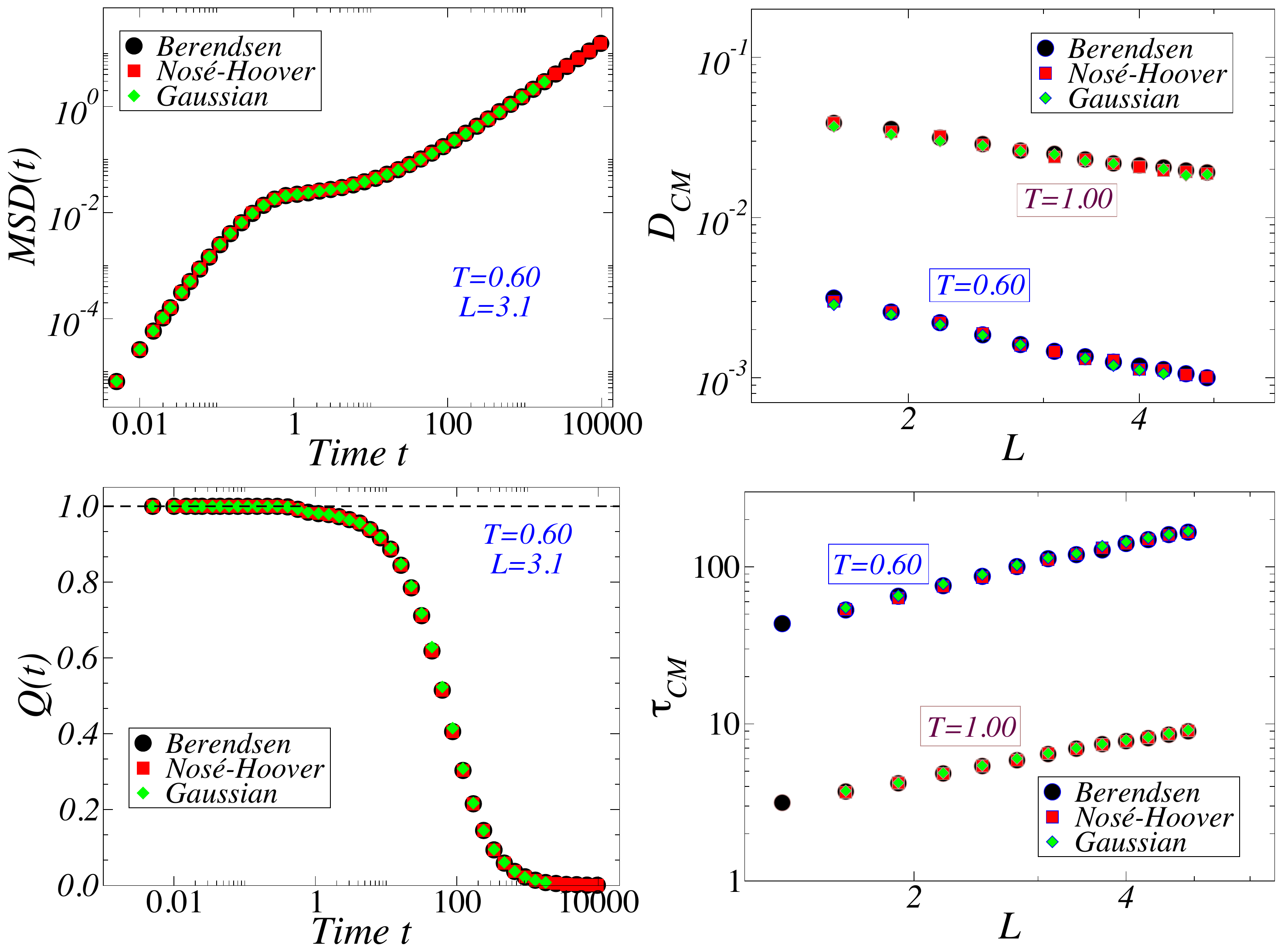}
	\caption{The time evolution of MSD using various thermostat-barostat combinations is shown in the top-left panel. The plots are for rod of length $L=3.1$ immersed in 3dKA model system at temperature, $T=0.60$. Here \textit{Berendsen} refers to Berendsen thermostat-barostat, \textit{Nos\'{e}-Hoover} refers to three-chain Nos\'{e}-Hoover thermostat-barostat, and \textit{Gaussian} refers to Gaussian thermostat-Berendsen barostat. The rod length variation of diffusion constant obtained from the MSD's is plotted in the top-right panel for two different temperatures, namely $T=0.60$ and $T=1.00$. In the very similar way the overlap-correlation function, $Q(t)$ and the relaxation time and $\tau_{CM}(T,L)$ is plotted in the bottom panel. The equivalence in all of these quantities with different thermostat-barostat suggests the independence of the reported results on thermostats and barostats. }
	\label{Thermostat}
\end{figure}
\noindent{\bf 3dHP Model:}
The 3dHP model is a paradigmatic model to study jamming physics in the context
of dense, soft sphere systems and soft granular particles, including colloidal 
particles in experiments. It is a 50:50 binary mixture of harmonic spheres with
diameter ratio 1.4 and interacting via potential,
\begin{equation}
V_{\alpha\beta}(\textbf{r})=\epsilon\left[1-\left(\frac{r_{\alpha\beta}}
{\sigma_{\alpha\beta}}\right)\right]^2
\end{equation}
for $r_{\alpha\beta} < \sigma_{\alpha\beta}$ and $V_{\alpha\beta}=0$ 
otherwise, where $\sigma_{\alpha\beta}=\frac{(\sigma_\alpha+\sigma_\beta)}{2}
$ and $\epsilon=1$. Again as in 3dKA, its pressure is chosen to keep average 
number density, $\rho=0.82$ with N=5000, while the temperature is varied 
between $\{T=0.0050,0.0090\}$\\
\noindent{\bf Rods:}
The embedded probe rods in the glass forming liquids are made of variable 
number of spheres, $N_b$, center of each sphere is separated from the consecutive 
sphere by a fixed distance of $0.3 \sigma_{AA}$ for 3dKA model and $0.42 \sigma_{AA}$
for 3dHP model. Each sphere of the rod have a same mass and interacts via same potential
as any other sphere in the system. The interaction parameters are spheres in the rod are, 
$\sigma_{RodSphere}=1$, $\epsilon_{RodSphere}=1$, $\sigma_{RodRod}=1$ and 
$\epsilon_{RodRod}=\frac{1}{2}$ same for all models. The rod length is defined as 
$L=0.3*(N_b-1)+1.0$ for 3dKA and 2dmKA models and $L=0.42*(N_b-1)+1.0$ for 3dHP model.

We have done NPT simulations for the models stated. 
Equation of motion for translational dynamics of spheres and center of mass (CoM) of rods is 
integrated by usual Leap-Frog integrator. The dynamics of rod's orientation vector ($
\hat{\textbf{u}}$) is also integrated by the same integrator but strictly following the methods 
illustrated in \cite{Rapaport2004, Allen}. The temperature and pressure in the simulation is 
controlled by Berendsen thermostat and barostat ~\cite{Allen}. A different thermostat does not 
change the results qualitatively. We reached this conclusion by checking our results with 
following different thermostat-barostat pairs,  
(a) three chain Nose-Hoover thermostat-barostat,
(b) Gaussian thermostat Berendsen barostat,
(c) Berendsen thermostat-barostat.
For consistency between the schemes, we have used same time damping constants for both Berendsen and Nose-Hoover ,viz $\tau_T=0.2$ and $\tau_P=2.0$ in reduced units. We have performed a consistency check on various quantities relevant to the results reported with above mentioned thermostat-barostat schemes, and is shown in Fig.\ref{Thermostat}. More details about the various thermostats and barostats can be found in Ref.\cite{Allen}.	 

\bibliography{Supli}